\documentclass[brochure,12pt]{bourbaki}
\usepackage[matrix,arrow]{xy}
\usepackage{graphicx,amssymb,amsfonts,amsmath,footnote}
\usepackage[francais]{babel}
\usepackage[applemac]{inputenc}

\usepackage[T1]{fontenc}

\date{Mars 2014}
\bbkannee{66\`eme ann\'ee, 2013-2014}
\bbknumero{1083}

\newcommand{\supess}{\operatorname{supess}}

\title{De Newton à Boltzmann et Einstein :\\ Validation des modèles cinétiques et de diffusion}
\subtitle{d'après T. Bodineau, I. Gallagher, L. Saint-Raymond, B. Texier}

\author{Fran\c cois GOLSE}
\address{\'Ecole polytechnique\\ Centre de mathématiques Laurent Schwartz\\ F--91128 Palaiseau Cedex}
\email{golse@math.polytechnique.fr}

\begin{document}
\maketitle


\noindent{\bf Introduction}


La théorie cinétique des gaz remonte aux travaux de Maxwell \cite{Max1,Max2} et de Boltzmann \cite{Boltz}. Toutefois, ni Maxwell ni Boltzmann n'ont véritablement cherché à établir la théorie cinétique des gaz comme limite des
équations de la mécanique classique écrites pour chaque molécule de gaz. C'est pourquoi le statut de la théorie cinétique des gaz est resté assez longtemps ambigu: était-ce une conséquence de la mécanique newtonienne, ou 
bien, au contraire, une théorie physique distincte qui, tout comme la thermodynamique, ne pouvait se déduire du principe fondamental de la dynamique? Cette question était d'autant plus troublante que les raisonnements utilisés 
par Maxwell et Boltzmann pour établir l'équation connue aujourd'hui sous le nom d'\textit{équation de Boltzmann} pouvaient se ramener à des considérations très simples de dynamique (conservation de l'impulsion et de l'énergie
cinétique des molécules de gaz avant et après collision) et à des arguments élémentaires de nature statistique. En même temps qu'il écrivait l'équation qui porte son nom, Boltzmann établissait en 1872 le \textit{théorème H} (cf.
(\ref{ThmH}) ci-dessous), une propriété mathématique liée aux symétries de l'équation de Boltzmann, analogue au second principe de la thermodynamique (voir \cite{Boltz}, partie I, sections 5, 6 et 8). Qu'un énoncé analogue à la 
croissance de l'entropie puisse être obtenu à partir des principes fondamentaux de la dynamique semblait contradictoire avec le caractère \textit{réversible} des équations de la mécanique (voir (\ref{RevMec}) ci-dessous), et surtout 
avec le théorème de récurrence de Poincaré, paru en 1890. Cette contradiction fut à l'origine d'une controverse scientifique entre Boltzmann, Loschmidt, Poincaré, Zermelo, dont on trouvera une description détaillée dans \cite{CeLB}. 
Lors du Congrès international des mathématiciens de 1900 à Paris, Hilbert pose le problème de l'{\og axiomatisation de la physique\fg}, et cite l'exemple suivant: {\og{\it le Livre de M. Boltzmann sur les Principes de la Mécanique nous incite 
à établir et à discuter au point de vue mathématique d’une manière complète et rigoureuse les méthodes basées sur l’idée de passage à la limite, et qui de la conception atomique nous conduisent aux lois du mouvement des continua.}\fg}
Près d'un demi-siècle après, Grad \cite{Grad} réussit à identifier un régime asymptotique dans lequel l'équation de Boltzmann pourrait être démontrée par passage à la limite à partir des équations de Newton de la mécanique classique. 
Finalement, Lanford proposa un schéma de preuve précis \cite{La} en 1975, établissant la validité de l'équation de Boltzmann sur un intervalle de temps court à partir des équations de la mécanique classique. Malgré tout l'article \cite{La}, 
ainsi que les diverses présentations du théorème de Lanford qui l'ont suivi, laissaient de côté la vérification rigoureuse de nombreux points techniques, vérification effectuée pour la première fois dans \cite{GSRT}. 

Après, il restait, pour compléter le programme esquissé par Hilbert, à obtenir à partir des équations de Newton une équation de la mécanique des milieux continus. Comme on l'a dit plus haut, le théorème de Lanford ne garantit la validité 
de l'équation de Boltzmann comme conséquence des équations de Newton de la mécanique classique que sur des intervalles de temps courts. On connaît d'autre part les régimes asymptotiques permettant d'établir rigoureusement la 
plupart des équations de la mécanique des fluides à partir de l'équation de Boltzmann (voir par exemple \cite{Vill}). Or ces régimes asymptotiques nécessitent de démontrer la validité de la théorie cinétique sur des intervalles de temps 
considérablement plus longs que ceux obtenus par Lanford. En restreignant leur étude au cas de l'équation de Boltzmann linéaire, Bodineau, Gallagher et Saint-Raymond ont réussi à établir l'équation de Boltzmann linéaire sur des 
intervalles de temps tendant vers l'infini avec le nombre de particules \cite{BGSR}. Ce résultat remarquable leur permet ensuite d'obtenir l'équation de diffusion (c'est-à-dire l'équation de la chaleur) --- et par voie de conséquence le 
mouvement brownien --- comme limite d'une dynamique {\it déterministe} de particules identiques en interaction.


\section{Présentation des modèles}


Commençons par donner un aperçu des différents modèles de dynamique particulaire dont il sera question dans cet exposé. Dans toute la suite, on supposera que la position des particules considérées varie dans l'espace euclidien 
$\mathbf{R}^3$ ou dans un tore plat de dimension $3$.

\subsection{Modèle n$^\circ$~1: les équations de Newton} 


Ce modèle décrit l'évolution d'un gaz monoatomique de manière exacte au niveau moléculaire. On considère que le gaz est un système de $N$ molécules sphériques de rayon $r$ et de même masse. Supposons dans un premier 
temps que les molécules ne sont soumises à aucune force extérieure et n'interagissent qu'au cours de collisions élastiques, et écrivons le principe fondamental de la dynamique\footnote{En pratique $r$ est très petit, de sorte que 
l'on peut négliger le mouvement de rotation de chaque molécule autour de son centre de gravité.} pour chaque molécule. Notant $x_k(t)\in\mathbf{R}^3$ et $v_k(t)\in\mathbf{R}^3$ pour $k=1,\ldots,N$ la position et la vitesse de la 
$k$-ième particule à l'instant $t\in\mathbf{R}$, on a donc
\begin{equation}\label{Newton}
\frac{dx_k}{dt}(t)=v_k(t)\,,\quad\frac{dv_k}{dt}(t)=0\,,\qquad\hbox{ si }|x_k(t)-x_l(t)|>2r\hbox{ pour tout }k\not=l\,.
\end{equation}
Au cours d'une collision entre la $k$-ième et la $l$-ième molécule à un instant $t^*$, les positions de ces molécules varient continûment en temps, c'est-à-dire que
\begin{equation}\label{Collxk}
x_k(t^*+0)=x_k(t^*-0)\,,\quad x_l(t^*+0)=x_l(t^*-0)\,,
\end{equation}
tandis que leurs vitesses varient de façon discontinue comme suit:
\begin{equation}\label{Collvkvj}
\begin{aligned}
v_k(t^*+0)=v_k(t^*-0)-((v_k(t^*-0)-v_l(t^*-0))\cdot n_{kl}(t^*))\,n_{kl}(t^*)\,,
\\
v_l(t^*+0)=v_l(t^*-0)+((v_k(t^*-0)-v_l(t^*-0))\cdot n_{kl}(t^*))\,n_{kl}(t^*)\,,
\end{aligned}
\end{equation}
en notant $n_{kl}(t^*):=(x_k(t^*\pm0)-x_l(t^*\pm0))/2r$. On notera dans la suite de cet exposé
$$
\Omega^r_N:=\{(x_1,\ldots,x_N)\in(\mathbf{R}^3)^N\hbox{ t.q. }|x_k-x_l|>2r\hbox{ pour tous }k,l=1,\ldots,N\,,\,\,k\not=l\}
$$
--- il s'agit de l'ensemble des positions physiquement admissibles pour les molécules, qui ne peuvent s'interpénétrer --- et $\Gamma^r_N=\Omega^r_N\times(\mathbf{R}^3)^N$. On suppose connues les positions et les vitesses 
de chaque molécule à l'instant initial $t=0$, soit
\begin{equation}\label{CondinNewton}
x_k(0)=x_k^{in}\,,\quad v_k(0)=v_k^{in}\,,\qquad\qquad k=1,\ldots,N
\end{equation}
avec $(x_1^{in},\ldots,x_N^{in},v_1^{in},\ldots,v_N^{in})\in\Gamma^r_N$, et on étudie le problème de Cauchy (\ref{Newton})-(\ref{Collxk})-(\ref{Collvkvj}) avec la condition initiale (\ref{CondinNewton}). Plus précisément, on 
cherche les solutions de ce problème de Cauchy $t\mapsto(x_1(t),\ldots,x_N(t),v_1(t),\ldots,v_N(t))$ à valeurs dans $\Gamma^r_N$.

Ce modèle semble être le plus précis que l'on puisse imaginer dans le cadre de la mécanique classique. On peut bien sûr penser que les interactions moléculaires dans un gaz sont plus complexes que des collisions élastiques
entre particules sphériques, mais là n'est pas l'essentiel.

En effet, si l'on veut utiliser ce modèle dans le cadre de la dynamique des gaz, il faut pouvoir traiter le cas d'un très grand nombre $N$ de particules. Typiquement, $N$ doit être de l'ordre du nombre d'Avogadro ($6.02\cdot 10^{23}$)
et $r$ est très petit (de l'ordre de $10^{-10}m$), ce qui rend la résolution du système (\ref{Newton})-(\ref{Collxk})-(\ref{Collvkvj}) impossible en pratique. Et même si l'on pouvait résoudre numériquement ce système avec la précision 
voulue, il resterait à en donner la condition initiale (\ref{CondinNewton}) avec la même précision. Or il est évidemment illusoire d'espérer connaître les positions et les vitesses instantanées de toutes les molécules d'un volume de gaz 
donné à un instant quelconque.

\subsection{Modèle n$^\circ$~2: l'équation de Boltzmann}


La théorie cinétique des gaz abandonne l'idée de déterminer la position et la vitesse de chaque molécule, tout en conservant la vision d'un gaz comme système de molécules identiques. Comme toutes les molécules de gaz sont 
identiques, l'idée de Maxwell \cite{Max2} et Boltzmann \cite{Boltz} est de décrire de manière statistique la dynamique d'une seule molécule {\og typique\fg}. L'état du gaz à tout instant $t$ est donné par sa \textit{fonction de distribution} 
$f\equiv f(t,x,v)\ge 0$, qui s'interprète comme suit. La fonction $(x,v)\mapsto f(t,x,v)$ est une densité de probabilité sur $\mathbf{R}^3\times\mathbf{R}^3$ à tout instant $t$, et pour tous $A,B\subset\mathbf{R}^3$ (mesurables),
$$
\frac{N_t(A,B)}{N}=\iint_{A\times B}f(t,x,v)dxdv\,,
$$
où $N$ est le nombre total de molécules, tandis que $N_t(A,B)$ désigne le nombre de molécules dont la vitesse appartient à $B$ et la position à $A$ à l'instant $t$. Par exemple, Maxwell \cite{Max1} établit que la fonction de 
distribution d'un gaz monoatomique porté à une température constante $T$ dans une enceinte cubique de volume unité est de la forme
\begin{equation}\label{Maxwell}
\mathcal{M}(t,x,v)=\left(\frac{m}{2\pi kT}\right)^{3/2}e^{-m|v|^2/2kT}\,,
\end{equation}
où $m$ est la masse atomique du gaz et $k$ la constante de Boltzmann. Mais cette formule ne vaut que pour des états d'équilibre. Pour décrire la dynamique d'un gaz en théorie cinétique, on écrit que sa fonction de distribution 
satisfait l'\textit{équation de Boltzmann}
\begin{equation}\label{EqBoltz}
(\partial_t+v\cdot\nabla_x)f=\mathcal{C}(f)\,,
\end{equation}
où $\mathcal{C}(f)$ est l'{\it intégrale de collision} de Boltzmann, qui s'exprime comme suit. Pour toute fonction $f\equiv f(v)$ indépendante de $t$ et $x$, continue et à décroissance rapide sur $\mathbf{R}^3$, 
\begin{equation}\label{IntColl}
\mathcal{C}(f)(v):=\gamma\iint_{\mathbf{R}^3\times\mathbf{S}^2}(f(v')f(v'_*)-f(v)f(v_*))((v-v_*)\cdot n)_+dv_*dn\,,
\end{equation}
où $\gamma>0$ est un paramètre tenant compte à la fois du nombre (très grand) de molécules par unité de volume et du diamètre moléculaire (très petit), et où
\begin{equation}\label{Collvv*v'v'*}
v'\equiv v'(v,v_*,n):=v-((v-v_*)\cdot n)\,n\,,\qquad v'_*\equiv v'_*(v,v_*,n):=v_*+((v-v_*)\cdot n)\,n\,.
\end{equation}
Pour toute fonction de distribution $f\equiv f(t,x,v)$ d\'ependant des variables $t,x,v$, continue et à décroissance rapide en la variable $v$, on d\'esigne par $\mathcal{C}(f)$ la fonction des variables $t,x,v$ d\'efinie par la formule 
$\mathcal{C}(f)(t,x,v):=\mathcal{C}(f(t,x,\cdot))(v)$. On a évidemment $\mathcal{C}(f)=\mathcal{C}_+(f)-\mathcal{C}_-(f)$, où
$$
\begin{aligned}
\mathcal{C}_+(f)(v):&=\gamma\iint_{\mathbf{R}^3\times\mathbf{S}^2}f(v')f(v'_*)((v-v_*)\cdot n)_+dv_*dn\,,
\\
\mathcal{C}_-(f)(v):&=\gamma f(v)\iint_{\mathbf{R}^3\times\mathbf{S}^2}f(v_*)((v-v_*)\cdot n)_+dv_*dn\,,
\end{aligned}
$$
nommés respectivement termes de gain et de perte dans l'intégrale de collision. Le terme de perte mesure le taux de disparition dans la population des particules de vitesse $v$ du fait des collisions avec d'autres particules (de 
vitesse quelconque $v_*$). Le terme de gain mesure le taux de création dans la population des particules de vitesse $v$ comme résultat d'une collision entre une particule de vitesse $v'$ et d'une particule de vitesse $v'_*$ avant
collision. On trouvera une présentation plus complète de l'équation de Boltzmann dans \cite{CIP}.

\subsection{Modèle n$^\circ$~3: l'équation de Boltzmann linéaire}


Une variante du modèle précédent consiste à étudier l'évolution de la fonction de distribution d'un gaz très raréfié mélangé à un autre gaz dans un état d'équilibre à une température uniforme $T$. On supposera pour simplifier que 
les deux gaz ont des molécules de même masse et de même diamètre. L'hypothèse de raréfaction permet de supposer, en première approximation, que la dynamique du gaz raréfié ne déstabilisera pas de manière sensible l'état 
d'équilibre de l'autre gaz. La fonction de distribution du gaz à l'équilibre est donc donnée par la formule de Maxwell (\ref{Maxwell}). La fonction de distribution $f$ du gaz raréfié satisfait l'\textit{équation de Boltzmann linéaire}
\begin{equation}\label{EqBoltzLin}
(\partial_t+v\cdot\nabla_x)f=\mathcal{L}(f)\,,
\end{equation}
où
\begin{equation}\label{IntCollLin}
\mathcal{L}(f)(v):=\gamma\left(\tfrac{m}{2\pi kT}\right)^{3/2}\iint_{\mathbf{R}^3\times\mathbf{S}^2}\left(f(v')e^{-\frac{m|v'_*|^2}{2kT}}-f(v)e^{-\frac{m|v_*|^2}{2kT}}\right)((v-v_*)\cdot n)_+dv_*dn\,,
\end{equation}
pour toute fonction $f\equiv f(v)$ indépendante de $t$ et $x$, continue et à décroissance rapide sur $\mathbf{R}^3$. Comme ci-dessus, on convient de noter $\mathcal{L}(f)(t,x,v):=\mathcal{L}(f(t,x,\cdot))(v)$ pour toute fonction de 
distribution $f\equiv f(t,x,v)$ d\'ependant des variables $t,x,v$, continue et à décroissance rapide en la variable $v$. 

\subsection{Modèle n$^\circ$~4: l'équation de diffusion}


Mais on peut également envisager de décrire la dynamique d'un gaz raréfié mélangé à un gaz en équilibre au moyen d'une \textit{équation de diffusion}. Ce dernier modèle oublie la vitesse individuelle des molécules du gaz raréfié. 
Soit $\rho\equiv\rho(t,x)\ge 0$ la densité macroscopique du gaz raréfié, qui s'interprète comme suit: pour tout $A\subset\mathbf{R}^3$ (mesurable),
$$
\frac{N_t(A)}{N}=\int_{A}\rho(t,x)dx\,,
$$
où $N$ est le nombre total de molécules du gaz raréfié, tandis que $N_t(A)$ désigne le nombre de ces molécules dont la position appartient à $A$ à l'instant $t$. Alors $\rho$ est solution de l'équation de diffusion
$$
\partial_t\rho-D\Delta_x\rho=0\,,
$$
où $D$ est un réel positif qui dépend de la fonction de distribution du gaz à l'équilibre, laquelle est donnée par la formule de Maxwell (\ref{Maxwell}). Du point de vue mathématique, l'équation de diffusion n'est évidemment rien 
d'autre que l'équation de la chaleur, proposée par Fourier afin de décrire la conduction thermique. Dans le contexte de la dynamique de particules microscopiques, l'équation de diffusion fut utilisée par Einstein \cite{Ein} qui proposa une 
interprétation mécanique du coefficient de diffusion $D>0$, et en déduisit un moyen de déterminer le nombre d'Avogrado\footnote{La valeur trouvée par Einstein sera ensuite précisée par Perrin, qui en donne une valeur comprise entre
$6,5$ et $6,8\cdot 10^{23}$ (cf. le texte de présentation par Oseen des travaux de Perrin à l'occasion du Prix Nobel de physique en 1926 sur le site de la fondation Nobel).}.


\section{De l'équation de Boltzmann linéaire à l'équation de diffusion}


Il s'agit de montrer que le modèle n$^\circ$~4 peut être obtenu à partir du modèle n$^\circ$~3 dans une certaine limite asymptotique. Cette étape du programme est aujourd'hui bien comprise. Posons $\beta:=m/kT$ et notons
\begin{equation}\label{DefMb}
M_\beta(v):=(\beta/2\pi)^{3/2}e^{-\beta|v|^2/2}\,.
\end{equation}
Considérons l'opérateur intégral $L_\beta$, défini pour toute fonction $\phi\in C_b(\mathbf{R}^3)$ par
$$
L_\beta\phi(v):=\gamma\iint_{\mathbf{R}^3\times\mathbf{S}^2}\Big(\phi(v)-\phi(v-(v-v_*)\cdot nn)\Big)M_\beta(v_*)((v-v_*)\cdot n)_+dv_*dn\,.
$$
On trouvera une démonstration des énoncés ci-dessous dans \cite{BSS,Pa}.

\begin{prop} 
Pour tout $\beta>0$, l'opérateur $L_\beta$ est un opérateur de Fredholm auto-adjoint non borné sur $L^2(\mathbf{R}^3;M_\beta dv)$, de domaine $D(L_\beta)=L^2(\mathbf{R}^3;(1+|v|^2)M_\beta dv)$. De plus $L_\beta\ge 0$, et 
$\mathrm{Ker}(L_\beta)$ est le sous-espace de $L^2(\mathbf{R}^3;M_\beta dv)$ formé des fonctions p.p. constantes sur $\mathbf{R}^3$, c'est-à-dire que $\mathrm{Ker}(L_\beta)=\mathbf{R}$.
\end{prop}

L'alternative de Fredholm et le fait que
$$
\int_{\mathbf{R}^3}vM_\beta(v)dv=0
$$
montrent qu'il existe un unique champ de vecteurs $A\in D(L_\beta)^3$ tel que
$$
L_\beta A(v)=v\,,\quad\hbox{ et }\int_{\mathbf{R}^3}A(v)M_\beta(v)dv=0\,.
$$
De plus 
$$
\int_{\mathbf{R}^3}v\cdot A(v)M_\beta(v)dv>0\,.
$$

\smallskip
Pour tout $\lambda>0$, on notera dans toute la suite de cet exposé $\mathbf{T}^3_\lambda:=\mathbf{R}^3/\lambda\mathbf{Z}^3$.

\begin{theo}\label{T-ApproxDiff}
Posons 
\begin{equation}\label{CoeffDiff}
D:=\tfrac13\int_{\mathbf{R}^3}v\cdot A(v)M_\beta(v)dv\,.
\end{equation}
Pour tout $\rho^{in}\in C^4(\mathbf{T}^3)$, soit $\rho$ l'unique solution bornée du problème de Cauchy
$$
\partial_s\rho-D\Delta_y\rho=0\,,\quad y\in\mathbf{T}^3\,,\qquad\qquad \rho\big|_{s=0}=\rho^{in}\,.
$$
Soit d'autre part, pour tout $\lambda>1$, la solution $f_\lambda$ de l'équation de Boltzmann linéaire
$$
\left\{
\begin{aligned}
{}&\partial_tf_\lambda+v\cdot\nabla_xf_\lambda+L_\beta f_\lambda=0\,,\quad x\in\mathbf{T}^3_\lambda\,,\,\,v\in\mathbf{R}^3\,,
\\
&f_\lambda(0,x,v)=\rho^{in}(x/\lambda)\,,\quad x\in\mathbf{T}^3_\lambda\,,\,\,v\in\mathbf{R}^3\,.
\end{aligned}
\right.
$$
Alors, pour tout $T>0$, il existe $C_T>0$ tel que
$$
\sup_{0\le s\le T\atop y\in\mathbf{T}^3,v\in\mathbf{R}^3}|f_\lambda(\lambda^2s,\lambda y,v)-\rho(s,y)|\le C_T/\lambda\,.
$$
\end{theo}


\section{De la mécanique de Newton à l'équation de Boltzmann}


Il s'agit là de la partie la plus conceptuelle du programme. Une première difficulté évidente vient des différences formelles entre les équations de Newton et l'équation de Boltzmann. En effet, les premières forment un système d'équations
différentielles ordinaires posées dans un espace des phases de dimension $6N$, tandis que la second est une équation intégro-différentielle posée dans un espace des phases de dimension $6$. Dans cette section, nous allons mettre en 
place une procédure formelle expliquant comment relier les modèles n$^\circ$~1 et n$^\circ$~2. Cette procédure, due à Cercignani \cite{Ce}, donnera une idée précise du déroulement de la preuve, mais il restera de nombreuses difficultés 
techniques à lever.

\subsection{De la mécanique de Newton à l'équation de Liouville}


Puisque le membre de gauche de l'équation de Boltzmann fait intervenir l'opérateur aux dérivées partielles $\partial_t+v\cdot\nabla_x$, une première étape naturelle est de remplacer le système d'équations différentielles ordinaires
(\ref{Newton}) par une équation aux dérivées partielles d'ordre $1$. 

Commen\c cons par préciser la notion de solution des équations de la mécanique de Newton. On notera $m_N$ la mesure de Lebesgue sur $(\mathbf{R}^3)^N\times(\mathbf{R}^3)^N$.

\begin{prop}
Il existe $E\subset\overline{\Gamma^r_N}$ tel que $m_N(E)=0$ et vérifiant la propriété suivante: pour tout $(x^{in}_1,\ldots,x^{in}_N,v^{in}_1,\ldots,v^{in}_N)\in\overline{\Gamma^r_N}\setminus E$, le problème de Cauchy
(\ref{Newton})-(\ref{Collxk})-(\ref{Collvkvj})-(\ref{CondinNewton}) admet une unique solution
$$
t\mapsto(x_1(t),\ldots,x_N(t),v_1(t),\ldots,v_N(t))=:S_t^{N,r}(x^{in}_1,\ldots,x^{in}_N,v^{in}_1,\ldots,v^{in}_N)\,.
$$
définie pour tout $t\in\mathbf{R}$. Ceci définit $S^{N,r}_t$ comme groupe à $1$ paramètre de transformations sur $\overline{\Gamma^r_N}\setminus E$ pour tout $t\in\mathbf{R}$. Autrement dit, pour tout $t\in\mathbf{R}$, l'application 
$S^{N,r}_t$ envoie $\overline{\Gamma^r_N}\setminus E$ dans lui-même, et on a
$$
S^{N,r}_{t+s}(x^{in}_1,\ldots,x^{in}_N,v^{in}_1,\ldots,v^{in}_N)=S^{N,r}_{t}(S^{N,r}_{s}(x^{in}_1,\ldots,x^{in}_N,v^{in}_1,\ldots,v^{in}_N))
$$
pour tous $t,s\in\mathbf{R}$ et $(x^{in}_1,\ldots,x^{in}_N,v^{in}_1,\ldots,v^{in}_N)\in\overline{\Gamma^r_N}\setminus E$. D'autre part, la mesure $m_N$ est invariante sous l'action de $S^{N,r}_t$:
$$
m_N(S^{N,r}_t(A))=m_N(A)\,,\quad\hbox{ pour toute partie mesurable }A\subset\overline{\Gamma^r_N}\hbox{ et tout }t\in\mathbf{R}\,.
$$
\end{prop}

La définition d'une notion de dynamique pour un nombre infini de sphères dures est due à Alexander \cite{Alex}. L'ensemble $E$ contient en particulier toutes les positions et vitesses initiales conduisant en temps fini à une collision 
mettant en jeu $3$ particules ou plus. On remarquera en effet que les conditions (\ref{Collxk})-(\ref{Collvkvj}) permettent de poursuivre la dynamique seulement lorsque les collisions ne mettent en jeu que deux particules. Montrer que 
cet ensemble est $m_N$-négligeable n'est pas très difficile: voir la proposition 4.1.1 dans \cite{GSRT}. L'invariance de $m_N$ sous l'action de $S^{N,r}_t$ est  claire: sur chaque intervalle de temps où il n'y a pas de collisions, la vitesse 
de chaque particule reste constante et sa position subit une translation, tandis que, lors d'une collision entre $2$ particules, la règle de transformation des vitesses (\ref{Collvkvj}) avant et après collision définit une isométrie linéaire 
sur $\mathbf{R}^3\times\mathbf{R}^3$.

\smallskip
On déduit de la proposition ci-dessus que, pour toute fonction mesurable positive ou nulle $F^{in}$ définie $m_N$-p.p. sur $\overline{\Gamma^r_N}$, la formule
$$
\tilde S^{N,r}_tF^{in}(x_1,\ldots,x_N,v_1,\ldots,v_N):=F^{in}(S^{N,r}_{-t}(x_1,\ldots,x_N,v_1,\ldots,v_N))
$$
définit de manière unique $m_N$-p.p. sur $\overline{\Gamma^r_N}$ une fonction mesurable positive ou nulle $\tilde S^{N,r}_tF^{in}$  pour tout $t\in\mathbf{R}$. De plus cette fonction $\tilde S^{N,r}_tF^{in}$ vérifie la loi de conservation
\begin{equation}\label{LiouvThm}
\int_{\Gamma^r_N}\Phi(\tilde S^{N,r}_tF^{in})dm_N=\int_{\Gamma^r_N}\Phi(F^{in})dm_N
\end{equation}
pour toute fonction $\Phi:\,\mathbf{R}_+\to\mathbf{R}_+$ continue, par invariance de la mesure $m_N$ sous l'action du groupe à $1$ paramètre $S^{N,r}_t$. 

\smallskip
Lorsque $F^{in}$ est la \textit{fonction de distribution jointe} du système de $N$ particules à $t=0$, la fonction 
$$
F(t,x_1,\ldots,x_N,v_1,\ldots,v_N):=\tilde S^{N,r}_tF^{in}(x_1,\ldots,x_N,v_1,\ldots,v_N)
$$
construite ci-dessus fournit la fonction de distribution jointe du système de $N$ particules à tout instant $t\in\mathbf{R}$. Autrement dit, étant donnés $A_1,B_1,\ldots,A_N,B_N\subset\mathbf{R}^3$ mesurables, l'intégrale
$$
\int_{A_1\times\ldots\times A_N\times B_1\times\ldots\times B_N}\tilde S^{N,r}_tF^{in}(x_1,\ldots,x_N,v_1,\ldots,v_N)dm_N(x_1,\ldots,x_N,v_1,\ldots,v_N)
$$
donne la probabilité qu'à l'instant $t$, la particule n$^\circ$~1 soit localisée dans $A_1$ avec une vitesse appartenant à $B_1$, la particule n$^\circ$~2 localisée dans $A_2$ avec une vitesse appartenant à $B_2$,\dots, et la particule 
n$^\circ$~$N$ localisée dans $A_N$ avec une vitesse appartenant à $B_N$.

\smallskip
De plus, la fonction $F=\tilde S^{N,r}_tF^{in}$ est l'unique solution au sens des distributions de l'\textit{équation de Liouville}
\begin{equation}\label{Liouville}
\begin{aligned}
\left(\frac{\partial F}{\partial t}+\sum_{k=1}^Nv_k\cdot\nabla_{x_k}F\right)(t,x_1,\ldots,x_N,v_1,\ldots,v_N)=0\,,
\\
(x_1,\ldots,x_N,v_1,\ldots,v_N)\in\Gamma^r_N\,,\quad t\in\mathbf{R}\,,
\end{aligned}
\end{equation}
avec la condition aux limites sur $\partial\Gamma^r_N$
\begin{equation}\label{BCLiouville}
\begin{aligned}
F&(t,x_1,\ldots,x_N,v_1,\ldots,v_N)
\\
&\qquad=F(t,x_1,\ldots,x_N,v_1,\ldots,v_{k-1},v'_k,v_{k+1},\ldots,v_{l-1},v'_l,v_{l+1},\ldots,v_N)\,,
\\
&\qquad\quad\qquad\quad |x_k-x_l|=2r\hbox{ et }|x_i-x_j|>2r\hbox{ pour tous }\{i,j\}=\{k,l\}\,,
\end{aligned}
\end{equation}
où $(v'_k,v'_l)$ est défini à partir de $(v_k,v_l)$ par les relations (\ref{Collvkvj}), et la condition initiale
\begin{equation}\label{CILiouville}
F(0,x_1,\ldots,x_N,v_1,\ldots,v_N)=F^{in}(x_1,\ldots,x_N,v_1,\ldots,v_N)\,.
\end{equation}
On trouvera plus de détails sur ces différents points dans les sections 4.2 et 4.3 de \cite{GSRT}.

\subsection{La hiérarchie BBGKY}\label{SS-HierBBGKY}


L'équation de Liouville, tout comme l'équation de Boltzmann, fait intervenir l'opérateur différentiel $\partial_t+v\cdot\nabla_x$, mais avec une différence importante: l'équation de Liouville est posée sur $(\mathbf{R}^3\times\mathbf{R}^3)^N$
--- plus précisément sur $\Gamma^r_N$ --- qui est l'espace des phases à $N$ particules, tandis que l'équation de Boltzmann est posée sur $\mathbf{R}^3\times\mathbf{R}^3$ qui est l'espace des phases d'une seule particule ponctuelle.

Pour comprendre cette réduction de la dimension de l'espace des phases, il faut bien entendu utiliser le fait que les $N$ molécules de gaz sont indistingables. Cette nouvelle information se traduit par le fait que
\begin{equation}\label{UndistF}
F(t,x_1,\ldots,x_N,v_1,\ldots,v_N)=F(t,x_{\sigma(1)},\ldots,x_{\sigma(N)},v_{\sigma(1)},\ldots,v_{\sigma(N)})
\end{equation}
pour tout $t\in\mathbf{R}$, presque tout $(x_1,\ldots,x_N,v_1,\ldots,v_N)\in\Gamma^r_N$ et toute permutation $\sigma\in\mathfrak{S}_N$. Grâce à l'unicité de la solution du problème de Cauchy (\ref{Newton})-(\ref{Collxk})-(\ref{Collvkvj}) 
avec la donnée initiale (\ref{CondinNewton}), on vérifie que si la condition (\ref{UndistF}) est satisfaite à l'instant initial $t=0$, elle est propagée par le groupe $\tilde S^{N,r}_t$ et donc est satisfaite pour tout $t\in\mathbf{R}$. 

A partir de là, il devient naturel de considérer que les variables $x_1,v_1$ dans $F$ se rapportent non pas à une molécule particulière, mais à la {\og molécule de gaz typique\fg}, et de faire disparaître les autres variables de $F$ par intégration.

Dans la suite de cet exposé, nous utiliserons systématiquement les notations suivantes: $X_N:=(x_1,\ldots,x_N)$ et $V_N:=(v_1,\ldots,v_N)$, tandis que $X_{k,N}:=(x_k,\ldots,x_N)$ et $V_{k,N}:=(v_k,\ldots,v_N)$. Pour toute fonction
$F_N\equiv F_N(t,X_N,V_N)$ mesurable positive ou nulle définie p.p. sur $\mathbf{R}^{6N+1}$, on pose
\begin{equation}\label{Marginal}
F_{N:k}(t,X_k,V_k):=\int F_N(t,X_N,V_N)dX_{k+1,N}dV_{k+1,N}
\end{equation}
pour tout $k$ tel que $1\le k\le N$. On conviendra de poser $F_{N:k}\equiv 0$ pour $k>N$. Bien évidemment, lorsque $F_N$ satisfait la condition de symétrie (\ref{UndistF}), $F_{N:k}$ vérifie également la condition (\ref{UndistF}) --- pour
$k$ paires de variables $x_j,v_j$ seulement, au lieu de $N$. Lorsque $F_N(t,\cdot)$ est une densité de probabilité sur $\mathbf{R}^{6N}$, on désigne la fonction $F_{N:k}$ sous le nom de $k$-ième \textit{marginale} de $F_N$.

On va donc essayer d'écrire une équation gouvernant l'évolution de la $1$ère marginale de $F(t,\cdot):=\tilde S^{N,r}_tF^{in}$. Cette fonction n'étant définie que p.p. sur $\overline{\Gamma^r_N}$, on la prolonge p.p. dans $\mathbf{R}^{6N}$ 
en posant
\begin{equation}\label{DefFN}
F_N(t,X_N,V_N):=\left\{\begin{array}{ll}F(t,X_N,V_N)&\hbox{ si }X_N\in\Omega^r_N\\ 0&\hbox{ si }X_N\notin\Omega^r_N\end{array}\right.
\end{equation}
Puis on observe que la fonction $F_N$ ainsi définie vérifie 
\begin{equation}\label{LiouvilleDist}
\partial_tF_N+\sum_{k=1}^Nv_k\cdot\nabla_{x_k}F_N=\sum_{1\le i<j\le N}F_N\Big|_{\partial^+\Gamma^r_N}(v_j-v_i)\cdot n_{ij}\delta_{|x_i-x_j|=2r}
\end{equation}
au sens des distributions sur $\mathbf{R}^{6N+1}$ (utiliser la formule (II.3.1) de \cite{Schw}). Dans le membre de droite de cette égalité, la notation $F_N\Big|_{\partial^+\Gamma^r_N}$ désigne la valeur limite de $F$ lorsque le $N$-uplet 
de positions $X_N$ tend vers le bord $\partial\Gamma^r_N$ tout en restant dans $\Gamma^r_N$, c'est-à-dire que
$$
F_N\Big|_{\partial^+\Gamma^r_N}=\lim_{\epsilon\to 0^+}F\Big|_{\partial\Gamma^{r+\epsilon}_N}\,.
$$
La notation $\delta_{|x_i-x_j|=2r}$ désigne la distribution de simple couche de densité $1$ sur la partie de $\partial\Omega^r_N$ où $|x_i-x_j|=2r$, et on rappelle que $n_{ij}=\frac{x_j-x_i}{|x_j-x_i|}$.

En intégrant chaque membre de cette égalité par rapport aux variables $x_2,\ldots,x_N$ et $v_2,\ldots,v_N$, on trouve que $F_{N:1}$ doit vérifier l'égalité suivante:
\begin{equation}\label{EqFN1a}
\begin{aligned}
(\partial_t+&v_1\cdot\nabla_{x_1})F_{N:1}(t,x_1,v_1)
\\
&=(N-1)(2r)^2\int_{\mathbf{R}^3\times\mathbf{S}^2}F_{N:2}(t,x_1,x_1+2rn,v_1,v_2)(v_2-v_1)\cdot ndv_2dn\,.
\end{aligned}
\end{equation}

En effet, pour $2\le i<j\le N$, on déduit de la condition aux limites (\ref{BCLiouville}) que\footnote{Les formules suivantes constituent un abus de notation manifeste, puisque $\delta_{|x_i-x_j|=2r}$ est une mesure de Radon et non une 
fonction.}
$$
\int F_N(t,X_N,V_N)(v_j-v_i)\cdot n_{ij}\delta_{|x_i-x_j|=2r}dX_{2,N}dV_{2,N}=0\,.
$$

Considérons par exemple le cas où $i=2$ et $j=3$: en utilisant (\ref{BCLiouville}) on trouve que
$$
\begin{aligned}
\int F_N(t,X_N,V_N)(v_3-v_2)\cdot n_{23}\delta_{|x_2-x_3|=2r}dX_{2,N}dV_{2,N}&
\\
=
-\int F_N(t,X_N,v_1,v'_2,v'_3,V_{4,N})(v'_3-v'_2)\cdot n_{23}\delta_{|x_2-x_3|=2r}dX_{2,N}dv_2dv_3dV_{4,N}&
\\
=-\int F_N(t,X_N,V_N)(v_3-v_2)\cdot n_{23}\delta_{|x_2-x_3|=2r}dX_{2,N}dV_{2,N}=0&\,,
\end{aligned}
$$
en effectuant le changement de variables $(v_2,v_3)\mapsto(v'_2,v'_3)$. 

D'autre part, lorsque $i=1$ et $3\le j\le N$
$$
\begin{aligned}
\int F_N(t,X_N,V_N)(v_j-v_1)\cdot n_{1j}\delta_{|x_1-x_j|=2r}dX_{2,N}dV_{2,N}
\\
=
\int F_N(t,X_N,V_N)(v_2-v_1)\cdot n_{12}\delta_{|x_1-x_2|=2r}dX_{2,N}dV_{2,N}
\\
=
\int F_{N:2}(t,x_1,x_2,v_1,v_2)(v_2-v_1)\cdot n_{12}\delta_{|x_1-x_2|=2r}dx_2dv_2\,.
\end{aligned}
$$
La $1$ère égalité s'obtient par le changement de variables $(x_2,v_2,x_j,v_j)\mapsto(x_j,v_j,x_2,v_2)$ en utilisant la relation de symétrie (\ref{UndistF}), tandis que la $2$ème s'obtient par intégration en les variables $x_3,\ldots,x_N$
et $v_3,\ldots,v_N$. Cette dernière intégrale est transformée en celle figurant au membre de droite de (\ref{EqFN1a}) en utilisant des coordonnées sphériques, de sorte que $x_2$ devient $x_1+2rn$ tandis que la mesure 
$\delta_{|x_1-x_2|=2r}dx_2$ devient $(2r)^2dn$.

Ensuite, on décompose l'intégrale au membre de droite de (\ref{EqFN1a}) sous la forme
$$
\begin{aligned}
\int_{\mathbf{R}^3\times\mathbf{S}^2}F_{N:2}(t,x_1,x_1+2rn,v_1,v_2)(v_2-v_1)\cdot ndv_2dn&
\\
=
\int_{(v_2-v_1)\cdot n>0}F_{N:2}(t,x_1,x_1+2rn,v_1,v_2)(v_2-v_1)\cdot ndv_2dn&
\\
+
\int_{(v_2-v_1)\cdot n<0}F_{N:2}(t,x_1,x_1+2rn,v_1,v_2)(v_2-v_1)\cdot ndv_2dn&\,.
\end{aligned}
$$
On utilise alors la condition aux limites (\ref{BCLiouville}) pour remplacer $F_{N:2}(t,x_1,x_1+2rn,v_1,v_2)$ par $F_{N:2}(t,x_1,x_1+2rn,v'_1,v'_2)$ dans la $1$ère intégrale au membre de droite de l'égalité ci-dessus, sachant que les
vitesses $v'_1,v'_2$ sont données en fonction de $v_1,v_2$ par la relation (\ref{Collvv*v'v'*}), où $(v_1,v'_1)$ joue le rôle de $(v,v')$, et $(v_2,v'_2)$ celui de $(v_*,v'_*)$. Enfin, on change $n$ en $-n$ dans la $2$ème intégrale au 
membre de droite de cette même égalité. On aboutit alors à l'égalité suivante vérifiée par $F_{N:1}$:
\begin{equation}\label{EqFN1}
(\partial_t+v_1\cdot\nabla_{x_1})F_{N:1}=(N-1)(2r)^2\mathcal{C}_N^{12}(F_{N:2})
\end{equation}
où
\begin{equation}\label{DefCN12}
\begin{aligned}
\mathcal{C}_N^{12}(F_{N:2})(t,x_1,v_1):=\int_{\mathbf{R}^3\times\mathbf{S}^2}&(F_{N:2}(t,x_1,x_1+2rn,T_{12}[n](v_1,v_2))
\\
&-F_{N:2}(t,x_1,x_1-2rn,v_1,v_2))((v_2-v_1)\cdot n)_+dv_2dn
\end{aligned}
\end{equation}
et
$$
T_{12}[n](v_1,v_2):=(v_1-((v_1-v_2)\cdot n)n,v_2-((v_2-v_1)\cdot n)n)\,.
$$
Cette égalité n'est évidemment pas sans évoquer l'équation de Boltzmann: on remarquera en particulier que les manipulations ci-dessus ont fait apparaître le terme $\mathcal{C}_N^{12}(F_{N:2})$ qui ressemble à l'intégrale de collision
de Boltzmann. Malheureusement ce n'est pas à proprement parler une équation pour $F_{N:1}$, puisque le terme de collision $\mathcal{C}_N^{12}(F_{N:2})$ fait intervenir $F_{N:2}$, qui n'est pas en général connue en fonction de 
$F_{N:1}$.

On cherche donc une équation permettant de déterminer $F_{N:2}$, en intégrant chaque membre de l'équation de Liouville (\ref{Liouville}) par rapport aux variables $x_3,\ldots,x_N,v_3,\ldots,v_N$. En procédant comme pour $F_{N:1}$,
on aboutit ainsi à une équation vérifiée par $F_{N:2}$ mettant en jeu $F_{N:3}$. 

Plus généralement, en appliquant le même raisonnement pour $k=3,\ldots,N-1$, on arrive à la suite d'équations indexées par $k=1,\ldots,N$:
\begin{equation}\label{EqFNk}
\begin{aligned}
\partial_tF_{N:k}+\sum_{j=1}^kv_j\cdot\nabla_{x_j}F_{N:k}=(N-k)(2r)^2\sum_{j=1}^k\mathcal{C}_N^{j,k+1}(F_{N:k+1})&
\\
+\sum_{1\le i<j\le k}F_{N:k}\Big|_{\partial\Gamma^r_k}(v_j-v_i)\cdot n_{ij}\,\delta_{|x_i-x_j|=2r}&\quad 1\le k\le N-1\,,
\end{aligned}
\end{equation}
où
$$
\begin{aligned}
\mathcal{C}_N^{j,k+1}(F_{N:k+1})(t,X_k,V_k):=\int_{\mathbf{R}^3\times\mathbf{S}^2}(F_{N:k+1}(t,X_k,x_j+2rn,T_{j,k+1}[n]V_{k+1})&
\\
-F_{N:k+1}(t,X_k,x_j-2rn,V_{k+1}))((v_{k+1}-v_j)\cdot n)_+dv_{k+1}dn&\,,
\end{aligned}
$$
et où
$$
T_{j,k+1}[n](V_{k+1}):=(v_1,\ldots,v_{j-1},v'_j,v_{j+1},\ldots,v_k,v'_{k+1})\,,
$$
avec
$$
v'_j=v_j-((v_{k+1}-v_j)\cdot n)n\,,\qquad v'_{k+1}=v_{k+1}-((v_{k+1}-v_j)\cdot n)n\,.
$$
Enfin, l'équation vérifiée par $F_{N:N}=F_N$ coïncide avec l'équation de Liouville (\ref{LiouvilleDist}) au sens des distributions.

La suite d'équations (\ref{EqFNk}) est connue sous le nom de \textit{hiérarchie BBGKY}\footnote{Les lettres BBGKY sont les initiales de N.N. Bogoliubov, M. Born, H.S. Green, J.G. Kirkwood et J. Yvon qui ont obtenu ces hiérarchies 
d'équations dans des contextes divers de la mécanique statistique hors équilibre: voir \cite{Bo,BoGr,Kir,Yv}.}. La hiérarchie BBGKY est déduite, comme on vient de le voir, de l'équation de Liouville vérifiée par $F_N$ au sens des 
distributions. Mais d'autre part la dernière équation de la hiérarchie BBGKY n'est rien d'autre que l'équation de Liouville (\ref{LiouvilleDist}) vérifiée par $F_N$ au sens des distributions. Donc la hiérarchie BBGKY est exactement 
équivalente à l'équation de Liouville (\ref{Liouville})-(\ref{BCLiouville}) pour le système de $N$-particules. Par conséquent, la hiérarchie BBGKY est aussi équivalente au système différentiel (\ref{Newton})-(\ref{Collxk})-(\ref{Collvkvj}) 
des équations de Newton décrivant le mouvement de $N$ particules sphériques identiques n'interagissant qu'au cours de collisions élastiques. En effet, ce système différentiel est précisément celui définissant les courbes 
caractéristiques le long desquelles les solutions de l'équation de Liouville sont constantes. 

On pourrait donc croire que l'étape consistant à passer de l'équation de Liouville à la hiérarchie BBGKY est inutile, puisque l'une et l'autre contiennent exactement autant d'information sur $F_N$. Pourtant, comme on va le voir, 
c'est l'examen de la hiérarchie BBGKY qui va nous donner la clef de deux étapes cruciales permettant d'aboutir finalement à l'équation de Boltzmann.

Traditionnellement, la hiérarchie BBGKY est présentée sans faire appel au formalisme des distributions: voir \cite{Ce,CIP} ou encore la section 4.3 dans \cite{GSRT}. Dans la pratique, on utilise surtout la formule intégrale (4.3.6)
ou (4.3.8) de \cite{GSRT}. La présence du dernier terme au membre de droite de (\ref{EqFNk}) permet de ne pas oublier que la $k$-ième équation de la hiérarchie BBGKY est posée sur $\Gamma^r_k$. Ce point est fondamental,
et se trouve à l'origine des recollisions, qui constituent une des principales difficultés à surmonter pour établir l'équation de Boltzmann à partir des équations de Newton.

\subsection{La loi d'échelle de Boltzmann-Grad}


Nous avons déjà remarqué la ressemblance frappante entre la première équation de la hiérarchie BBGKY (\ref{EqFN1}) et l'équation de Boltzmann (\ref{EqBoltz}). Si l'on considère simultanément l'intégrale des collisions de
Boltzmann (\ref{IntColl}) et le terme analogue $\mathcal{C}_N^{12}(F_{N:2})$ au membre de droite de (\ref{EqFN1}), dont la définition est donnée dans (\ref{DefCN12}), il est logique de s'intéresser au régime asymptotique
\begin{equation}\label{BGrad}
N\to+\infty\,,\quad r\to 0^+\,,\quad N(2r)^2=\gamma\,.
\end{equation}
Cette hypothèse d'échelle porte le nom de Boltzmann-Grad, car il semble que Grad ait été le premier à formuler de façon précise un régime asymptotique permettant d'établir l'équation de Boltzmann à partir des équations 
de Newton dans son article \cite{Grad}.

\subsection{La hiérarchie de Boltzmann}\label{SS-HierBoltz}


Supposons désormais que $N\to+\infty$ et que $r$ est lié à $N$ par la relation (\ref{BGrad}), c'est-à-dire que $r=r(N):=\tfrac12\sqrt{\gamma/N}$. Supposons en outre que 
$$
F_{N:k}\to F_k\quad\hbox{ pour tout }k\ge 1\hbox{ lorsque }N\to+\infty
$$
en un sens à préciser. En supposant que cette convergence a lieu au moins au sens des distributions dans $\mathbf{R}_+^*\times(\mathbf{R}^3)^k\times(\mathbf{R}^3)^k$, on peut passer à la limite dans le membre de gauche 
de (\ref{EqFNk}) pour trouver que
$$
\partial_tF_{N:k}+\sum_{j=1}^kv_j\cdot\nabla_{x_j}F_{N:k}\to\partial_tF_k+\sum_{j=1}^kv_j\cdot\nabla_{x_j}F_k
$$
au sens des distributions dans $\mathbf{R}_+^*\times(\mathbf{R}^3)^k\times(\mathbf{R}^3)^k$. 

D'autre part, on suppose que la convergence ci-dessus a lieu dans un sens suffisamment fort pour assurer que
$$
\mathcal{C}_N^{j,k+1}(F_{N:k+1})\to\mathcal{C}^{j,k+1}(F_{k+1})\quad\hbox{ lorsque }N\to+\infty
$$
au moins au sens des distributions dans $\mathbf{R}_+^*\times(\mathbf{R}^3)^k\times(\mathbf{R}^3)^k$ pour tout $j=1,\ldots,k$, où
$$
\begin{aligned}
\mathcal{C}^{j,k+1}(F_{k+1}):=\int_{\mathbf{R}^3\times\mathbf{S}^2}(&F_{k+1}(t,X_k,x_j,T_{j,k+1}[n]V_{k+1})
\\
&-F_{k+1}(t,X_k,x_j,V_{k+1}))((v_{k+1}-v_j)\cdot n)_+dv_{k+1}dn\,.
\end{aligned}
$$

Enfin, le dernier terme figurant au membre de droite de (\ref{EqFNk}) vérifie
\begin{equation}\label{RecollFNk}
\sum_{1\le i<j\le k}F_{N:k}\Big|_{\partial\Gamma^r_k}(v_j-v_i)\cdot n_{ij}\,\delta_{|x_i-x_j|=2r}\to 0
\end{equation}
au sens des distributions dans $\mathbf{R}_+^*\times(\mathbf{R}^3)^k\times(\mathbf{R}^3)^k$ pour tout $k\ge 1$ lorsque $N\to+\infty$, sous l'hypothèse que
$$
M_k:=\sup_{1\le i\le N\atop N\ge 1}\sup_{(t,X_k,V_k)\in\mathbf{R}^{6k+1}}\left(|V_k|F_{N:k}(t,X_k,V_k)\right)<+\infty
$$
pour tout $k\ge 1$. En effet, un calcul élémentaire montre que
$$
\begin{aligned}
\int\left|\sum_{1\le i<j\le k}F_{N:k}\Big|_{\partial\Gamma^r_k}(v_j-v_i)\cdot n_{ij}\,\delta_{|x_i-x_j|=2r}\right|dX_kdV_k
\\
\le 2k(k-1)M_k\tfrac43\pi (2r)^3\to 0
\end{aligned}
$$
lorsque $N\to+\infty$.

Par conséquent, la suite $(F_k)_{k\ge 1}$ obtenue comme limite de $F_{N:k}$ lorsque $N\to+\infty$ à $k$ fixé est solution de la hiérarchie \textit{infinie} d'équations
\begin{equation}\label{EqFk}
\partial_tF_k+\sum_{j=1}^kv_j\cdot\nabla_{x_j}F_k=\gamma\sum_{j=1}^k\mathcal{C}^{j,k+1}(F_{k+1})\,,\quad k\ge 1\,.
\end{equation}
Cette hiérarchie infinie d'équations est connue sous le nom de \textit{hiérarchie de Boltzmann} (voir \cite{CIP}, \cite{GSRT} section 4.4).

Evidemment, la hiérarchie BBGKY et la hiérarchie de Boltzmann se ressemblent beaucoup. Toutefois cette ressemblance formelle est relativement trompeuse, et il existe d'importantes différences entre ces deux hiérarchies. Pour 
commencer, la hiérarchie BBGKY ne comporte qu'un nombre \textit{fini} d'équations --- nombre certes très grand, puisqu'il s'agit de $N$ qui est le nombre total de particules --- tandis que la hiérarchie de Boltzmann, elle, est formée 
d'une infinité d'équations. 

Une autre différence est la présence des termes de la forme (\ref{RecollFNk}) au membre de droite de chaque équation (\ref{EqFNk}) de la hiérarchie BBGKY. Traditionnellement, la hiérarchie BBGKY est présentée sans utiliser le 
formalisme des distributions, et l'équation (\ref{EqFNk}) comme une équation vérifiée par les restrictions $F_{N:k}\Big|_{\Gamma^{r(N)}_k}$, équation posée sur $\mathbf{R}_+\times\Gamma^{r(N)}_k$ pour $2\le k\le N$, et complétée 
par les conditions aux limites déduites de (\ref{BCLiouville}) --- dont le terme (\ref{RecollFNk}) est, comme on l'a remarqué plus haut, la traduction dans le formalisme des distributions. Au contraire, chaque équation (\ref{EqFk}) de la 
hiérarchie de Boltzmann est posée sur $\mathbf{R}_+\times\mathbf{R}^{6k}$, et ne comporte donc aucun terme additionnel analogue à (\ref{RecollFNk}).

Une dernière différence entre la hiérarchie de Boltzmann et la hiérarchie BBGKY tient au fait que les intégrales de collisions dans celle-ci sont \textit{délocalisées} --- autrement dit l'opération de moyenne sur le vecteur unitaire $n$ 
dans $\mathcal{C}^{j,k+1}_N(F_{N:k+1})$ met en jeu la dépendance de $F_{N:k+1}$ à la fois par rapport à sa $k+1$-ème variable d'espace et par rapport à ses $j$-ième et $k+1$-ième variables de vitesse, tandis que l'intégrale de 
collision $\mathcal{C}^{j,k+1}(F_{k+1})$ ne fait intervenir que la dépendance de $F_{k+1}$ par rapport aux $j$-ième et $k+1$-ième variables de vitesse, les positions de la $j$-ième et de la $k+1$-ième particules étant confondues.

\subsection{De la hiérarchie de Boltzmann à l'équation de Boltzmann non linéaire}\label{SS-HierBoltzEqBoltz}


Il ne reste plus qu'à expliquer comment l'équation de Boltzmann peut être déduite de la hiérarchie de Boltzmann. En fait, l'équation et la hiérarchie de Boltzmann sont reliées par la propriété remarquable suivante.

Soit $f$ une solution de l'équation de Boltzmann (\ref{EqBoltz}). Posons, pour tout $k\ge 1$
\begin{equation}\label{ftensk}
F_k(t,X_k,V_k)=\prod_{j=1}^kf(t,x_j,v_j)\,.
\end{equation}
Alors
$$
\frac{\partial F_k}{\partial t}+\sum_{j=1}^kv_j\cdot\nabla_{x_j}F_k(t,X_k,V_k)
=
\sum_{j=1}^k\gamma\mathcal{C}(f)(t,x_j,v_j)\prod_{l=1\atop l\not=j}^kf(t,x_l,v_l)
$$
pour tout $k\ge 1$. Un calcul trivial montre alors que
$$
\mathcal{C}(f)(t,x_j,v_j)\prod_{l=1\atop l\not=j}^kf(t,x_l,v_l)=\mathcal{C}^{j,k+1}(F_{k+1})(t,X_k,V_k)\,,
$$
de sorte que la suite $F_k$ construite à partir de la solution de l'équation de Boltzmann par la formule (\ref{ftensk}) est une solution de la hiérarchie de Boltzmann (\ref{EqFk}).

Par conséquent, si on sait démontrer 

\smallskip
\noindent
(a) que l'équation de Boltzmann admet une solution $f$ definie sur $[0,T]\times\mathbf{R}^3\times\mathbf{R}^3$, et 

\noindent
(b) que toute solution $(F_k)_{k\ge 1}$ de la hiérarchie de Boltzmann est déterminée de manière unique par sa valeur à l'instant $t=0$, on conclut que
$$
\hbox{si }F_k(0,X_k,V_k)=\prod_{j=1}^kf(0,x_j,v_j)\,,\quad\hbox{ alors }F_k(t,X_k,V_k)=\prod_{j=1}^kf(t,x_j,v_j)
$$
pour tout $t\in[0,T]$.

\smallskip
Si tel est le cas, et si les différents passages à la limite ci-dessus peuvent être justifiés, on aura ainsi montré que, pour tout $t\in[0,T]$,
\begin{equation}\label{ConvFN1}
F_{N:1}(t,x_1,v_1):=\int_{|x_i-x_j|>2r\atop1\le i<j\le N}F_N^{in}(S^{N,r}_{-t}(x_1,\ldots,x_N,v_1,\ldots,v_N))dX_{2,N}dV_{2,N}
\\
\to f(t,x_1,v_1)
\end{equation}
lorsque $N\to+\infty$ et pour $r$ défini par la loi d'échelle de Boltzmann-Grad (\ref{BGrad}), où la convergence ci-dessus a lieu en un sens restant à préciser. 

Il reste bien évidemment à vérifier la condition du (b) sur les données initiales pour la hiérarchie de Boltzmann. Soit $f^{in}=f\Big|_{t=0}$ la donnée initiale pour l'équation de Boltzmann; on ne peut évidemment pas prescrire
$$
F_N^{in}(t,X_N,V_N)=\prod_{j=1}^Nf^{in}(x_j,v_j)\,,
$$
puisque $F_N^{in}(t,X_N,V_N)=0$ pour tout $N$-uplet de positions $X_N\in(\mathbf{R}^3)^N$ tel qu'il existe deux indices $1\le i<j\le N$ pour lesquels $|x_i-x_j|<2r$. Ce point est d'ailleurs une différence supplémentaire entre la hiérarchie 
BBGKY et la hiérarchie de Boltzmann, pour laquelle la condition initiale
$$
F_k(0,X_k,V_k)=\prod_{j=1}^kf^{in}(x_j,v_j)\,,\quad  k\ge 1
$$
est évidemment admissible. En effet, la hiérarchie de Boltzmann est obtenue après passage à la limite de Boltzmann-Grad, donc pour $r=0$. 

La convergence (\ref{ConvFN1}) explique précisément comment l'équation de Boltzmann non linéaire, de solution $f$, est obtenue comme limite de la mécanique de Newton dans l'asymptotique de Boltzmann-Grad, puisque l'intégrale 
ci-dessus fait intervenir le groupe à un paramètre $S^{N,r}_{-t}$ engendré par les équations de Newton.

\subsection{De la mécanique de Newton à l'équation de Boltzmann linéaire}


Avant de donner un aperçu des méthodes mises en jeu dans la démonstration de la limite de Boltzmann-Grad --- c'est-à-dire dans l'obtention du modèle n$^\circ$~2 (l'équation de Boltzmann non linéaire) à partir du modèle n$^\circ$~1 
(les équations de Newton) --- expliquons comment le raisonnement formel décrit ci-dessus permet d'aboutir à l'équation de Boltzmann \textit{linéaire} (le modèle n$^\circ$~3). Précisons tout de suite qu'il ne s'agit pas de linéariser les 
divers modèles étudiés ci-dessus: en effet, l'équation de Liouville et les hiérarchies BBGKY et de Boltzmann qui en découlent sont toutes linéaires. La non linéarité de l'équation de Boltzmann ne provient en définitive que du choix de la
donnée initiale (\ref{CondInBBGKY}) et de la forme factorisée (\ref{ftensk}) qu'elle implique sur la solution de la hiérarchie de Boltzmann, une fois établi l'argument d'unicité (b) ci-dessus.

La première étape dans l'obtention de l'équation de Boltzmann linéaire à partir des équations de la mécanique de Newton consiste à considérer une particule marquée --- par exemple blanche --- évoluant au milieu d'un grand nombre
de particules noires, sachant que la particule blanche et les particules noires ont même rayon et même masse. Ces particules sont donc mécaniquement identiques, de sorte que les équations de Newton (\ref{Newton}) ainsi que 
les lois de collision (\ref{Collxk})-(\ref{Collvkvj}) sont les mêmes que celles considérées dans les sections précédentes. L'équation de Liouville (\ref{Liouville}) et les conditions aux limites (\ref{BCLiouville}), ainsi que la formulation de 
(\ref{Liouville})-(\ref{BCLiouville}) au sens des distributions, soit (\ref{LiouvilleDist}), demeurent inchangées.

En revanche, les particules considérées ne sont plus toutes indistingables, puisque l'une de ces particules est blanche et que toutes les autres sont noires. Convenons à partir de maintenant que $F_N(t,x_1,X_{2,N},v_1,V_{2,N})$ 
désigne la fonction de distribution du système formé de $N-1$ particules noires et d'$1$ particule blanche, les variables $x_1$ et $v_1$ désignant respectivement la position et la vitesse de la particule blanche. La relation de symétrie
(\ref{UndistF}) n'est donc plus satisfaite pour \textit{toute} permutation $\sigma\in\mathfrak{S}_N$, mais seulement pour celles fixant l'indice $1$. En désignant par $\mathfrak{S}_N^1$ le fixateur de $1$ dans $\mathfrak{S}_N$, on a donc
\begin{equation}\label{UndistLF}
F(t,x_1,\ldots,x_N,v_1,\ldots,v_N)=F(t,x_{\sigma(1)},\ldots,x_{\sigma(N)},v_{\sigma(1)},\ldots,v_{\sigma(N)})
\end{equation}
pour tout $t\in\mathbf{R}$, tout $(x_1,\ldots,x_N,v_1,\ldots,v_N)\in\Gamma^r_N$ et toute permutation $\sigma\in\mathfrak{S}_N^1$. Toujours grâce à l'unicité de la solution du problème de Cauchy (\ref{Newton})-(\ref{Collxk})-(\ref{Collvkvj}) 
avec la donnée initiale (\ref{CondinNewton}), on vérifie que si la condition (\ref{UndistLF}) est satisfaite à l'instant initial $t=0$, elle est propagée par le groupe $\tilde S^{N,r}_t$ et donc satisfaite pour tout $t\in\mathbf{R}$. 

A partir de là, on vérifie sans aucune difficulté que les arguments de symétrie utilisés dans la section \ref{SS-HierBBGKY} valent encore et montrent que les marginales $F_{N:k}$ sont encore solutions de la même hiérarchie BBGKY
(\ref{EqFNk}) que ci-dessus. Et donc, toujours sous l'hypothèse que $N$ et $r$ sont reliés par la loi d'échelle de Boltzmann-Grad (\ref{BGrad}), le même raisonnement qu'à la section \ref{SS-HierBoltz} montre que la suite $(F_k)_{k\ge 1}$ est 
solution de la même hiérarchie infinie d'équations (\ref{EqFk}), c'est-à-dire de la hiérarchie de Boltzmann.

Comme on l'a suggéré ci-dessus, c'est le choix de la donnée initiale dans l'équation de Liouville --- et donc dans la hiérarchie de Boltzmann (\ref{EqFk}) --- qui va faire toute la différence entre la dérivation de l'équation de Boltzmann
linéaire et celle de l'équation de Boltzmann non linéaire à partir de la mécanique de Newton. 

Dans la suite de cette section, on va supposer que les particules noires sont à l'équilibre. Ceci se traduit de la manière suivante: après passage à la limite de Boltzmann-Grad, la fonction de distribution des $N-1$ particules noires
correspondant aux indices $j=2,\ldots,N$ sera de la forme
$$
Y_N\prod_{j=2}^NM_\beta(v_j)\,,
$$
où $Y_N$ est un facteur de normalisation et où $M_\beta$ est la maxwellienne (\ref{DefMb}). Comme cette expression doit être une densité de probabilité sur l'espace des phases correspondant aux $N-1$ particules noires de rayon nul, 
il faut donc que l'espace des positions soit de mesure de Lebesgue finie. Dans toute la suite, on supposera donc que toutes les particules sont contenues dans $\mathbf{T}^3_\lambda:=\mathbf{R}^3/\lambda\mathbf{Z}^3$ de sorte que 
$Y_N:=\lambda^{3(1-N)}$. Quant à la loi d'échelle de Boltzmann-Grad (\ref{BGrad}), on va la modifier comme suit:
\begin{equation}\label{BGradLin}
N\to+\infty\,,\quad r\to 0^+\,,\quad\hbox{ et }N(2r)^2=\gamma\lambda^3\,.
\end{equation}

Tout d'abord, l'énoncé analogue au point (a) de la section précédente ne pose aucune difficulté dans le cas de l'équation de Boltzmann linéaire. L'observation qui va jouer, pour l'équation de Boltzmann linéaire, un rôle analogue à
celle de la section précédente est l'énoncé suivant.

\smallskip
Soit $f^{in}\equiv f^{in}(x,v)$, une densité de probabilité appartenant à $C_b(\mathbf{T}^3_\lambda\times\mathbf{R}^3)$ et soit $f$ la solution du problème de Cauchy pour l'équation de Boltzmann linéaire (\ref{EqBoltzLin}) v\'erifiant la
condition initiale $f\Big|_{t=0}=f^{in}$. Posons
$$
F_k(t,X_k,V_k)\!\!:=\!\!\lambda^{3(1-N)}f(t,x_1,v_1)\prod_{j=2}^kM_\beta(v_j)\,,\quad(x_j,v_j)\in\mathbf{T}^3_\lambda\times\mathbf{R}^3\,,\,\,1\le j\le k\,.
$$
Alors la suite $(F_k)_{k\ge 1}$ est solution de la hiérarchie de Boltzmann (\ref{EqFk}), avec donnée initiale
$$
F_k(0,X_k,V_k)=\lambda^{3(1-N)}f^{in}(x_1,v_1)\prod_{j=2}^kM_\beta(v_j)\,.
$$

En admettant que la hiérarchie de Boltzmann vérifie la propriété d'unicité (b) de la section précédente, on aboutit à l'énoncé suivant. Posons
$$
\begin{aligned}
F_N^{in}(X_N,V_N):=\mathcal{Z}_N^{-1}f^{in}(x_1,v_1)\prod_{j=2}^kM_\beta(v_j)\,,\quad |x_i-x_j|>2r\hbox{ pour }1\le i<j\le N\,,
\\
\quad(x_j,v_j)\in\mathbf{T}^3_\lambda\times\mathbf{R}^3\,,
\end{aligned}
$$
avec
$$
\mathcal{Z}_N:=\int_{(\mathbf{T}^3_\lambda)^N\times(\mathbf{R}^3)^N}f^{in}(x_1,v_1)\prod_{j=2}^kM_\beta(v_j)\prod_{1\le i<j\le N}\mathbf{1}_{|x_i-x_j|>2r}dX_NdV_N\,.
$$

Alors
$$
F_{N:1}(t,x_1,v_1):=\int_{|x_i-x_j|>2r\atop1\le i<j\le N}F_N^{in}(S^{N,r}_{-t}(x_1,\ldots,x_N,v_1,\ldots,v_N))dX_{2,N}dV_{2,N}\to f(t,x_1,v_1)
$$
où $f$ est la solution du problème de Cauchy pour l'équation de Boltzmann linéaire (\ref{EqBoltzLin}) avec condition initiale $f\Big|_{t=0}=f^{in}$. 

\section{Les principaux résultats}


Cette section regroupe les principaux énoncés de \cite{GSRT} et \cite{BGSR}.

\subsection{De la mécanique de Newton à l'équation de Boltzmann non linéaire}


Commen\c cons par le théorème de Lanford. Le résultat démontré dans \cite{GSRT} s'énonce comme suit.

\begin{theo}\label{T-Lanford}
Pour tous $\beta_0>0$ et $\mu_0\in\mathbf{R}$, il existe $T^*=T^*[\beta_0,\mu_0]>0$ vérifiant ce qui suit. Soit $f^{in}\equiv f^{in}(x,v)$ continue sur $\mathbf{R}^3\times\mathbf{R}^3$ vérifiant
$$
0\le f^{in}(x,v)\le e^{-\mu_0-\frac12\beta_0|v|^2}\hbox{ p.p. sur }\mathbf{R}^3\times\mathbf{R}^3\hbox{ et }\int_{\mathbf{R}^3\times\mathbf{R}^3}f^{in}(x,v)dxdv=1\,.
$$
Pour tout $N\ge 1$, on pose $r=\tfrac12\sqrt{\gamma/N}$ et
\begin{equation}\label{CondInBBGKY}
F_N^{in}(X_N,V_N):=\left(\int_{\Gamma^r_N}\prod_{k=1}^Nf^{in}(x_k,v_k)dX_NdV_N\right)^{-1}\prod_{k=1}^Nf^{in}(x_k,v_k)
\end{equation}
pour tout $(X_N,V_N)\in\Gamma^r_N$. Pour tout $t\in\mathbf{R}$, on définit
$$
F_N(t,X_N,V_N):=F_N^{in}(S^{N,r}_{-t}(X_N,V_N))\quad\hbox{ pour presque tout }(X_N,V_N)\in\Gamma^r_N\,,
$$
où on rappelle que $S^{N,r}_t$ est le groupe à un paramètre engendré par les équations de Newton (\ref{Newton})-(\ref{Collxk})-(\ref{Collvkvj}) pour un système de $N$ boules de rayon $r$ interagissant au cours de collisions élastiques. 
Alors, pour tout $\phi\in C_c(\mathbf{R}^3)$
$$
\int_{\mathbf{R}^3}F_{N:1}(t,x,v)\phi(v)dv\to\int_{\mathbf{R}^3}f(t,x,v)\phi(v)dv
$$
lorsque $N\to+\infty$ localement uniformément en $(t,x)\in[0,T^*[\times\mathbf{R}^3$, où $f$ est la solution de l'équation de Boltzmann (\ref{EqBoltz}) avec condition initiale $f\big|_{t=0}=f^{in}$.
\end{theo}

\smallskip
En réalité, le résultat obtenu est plus fort car il décrit la limite de toutes les marginales $F_{N:k}$ lorsque $N\to +\infty$: pour tout $k\ge 1$ et tout $\phi\in C_c((\mathbf{R}^3)^k)$
$$
\int_{(\mathbf{R}^3)^k}F_{N:k}(t,X_k,V_k)\phi(V_k)dV_k\to\int_{(\mathbf{R}^3)^k}\prod_{j=1}^kf(t,x_j,v_j)\phi(V_k)dV_k
$$
localement uniformément sur $\mathbf{R}_+\times\Omega^0_k$. Une suite de fonctions de $N$ variables vérifiant cette propriété est appelée \textit{suite chaotique}. La condition initiale $F^{in}_N$ que l'on utilise dans le théorème 
\ref{T-Lanford} est un exemple de suite chaotique; la stratégie de Lanford consiste à montrer que le caractère chaotique de la donnée initiale est propagé par le groupe à un paramètre $S^{N,r}_t$ engendré par les équations de Newton 
(\ref{Newton}) dans la limite de Boltzmann-Grad. La propagation du chaos se trouve au c\oe ur de très nombreux résultats de physique statistique: voir par exemple l'exposé \cite{Desv} sur le programme de Kac.

\smallskip
Le fait que la convergence ci-dessus n'ait lieu que sur un intervalle de temps très court est évidemment une sérieuse limitation. En utilisant l'effet dispersif de l'opérateur de transport $\partial_t+v\cdot\nabla_x$ décrit dans \cite{IS}, Illner 
et Pulvirenti ont réussi dans \cite{IP} à adapter l'argument de Lanford pour obtenir la limite de Boltzmann-Grad pour tout temps positif, mais en se limitant à des conditions initiales correspondant à une hypothèse de gaz {\og très raréfié\fg}
--- voir aussi la section 4.5 dans \cite{CIP}. Toutefois ce résultat ne permet pas, lui non plus, d'atteindre les régimes asymptotiques où les équations de la mécanique des fluides peuvent être déduites de l'équation de Boltzmann \cite{Vill}. 
Et bien que des solutions (certes en un sens très faible) de l'équation de Boltzmann aient été construites pour tout temps et pour toute donnée initiale de masse, d'énergie et d'entropie finie \cite{DiPL,Gé1}, les conditions sous lesquelles
on sait, à ce jour, déduire l'équation de Boltzmann des équations de la mécanique classique demeurent considérablement plus restrictives.

\subsection{De la mécanique de Newton à l'équation de Boltzmann linéaire}


Le résultat central de \cite{BGSR} s'énonce comme suit.

\begin{theo}\label{T-BGSR}
Soit $\phi_N^{in}\in C^1(\mathbf{T}^3_\lambda\times\mathbf{R}^3)$ vérifiant les conditions
$$
\frac1{\mu_N}\le\phi_N^{in}(x,v)\le\mu_N\,,\quad\hbox{ et }|\nabla\phi^{in}_N(x,v)|\le\Lambda_N\hbox{ pour tout }(x,v)\in\mathbf{T}^3_\lambda\times\mathbf{R}^3\,,
$$
où
$$
\mu_N=o(\sqrt{\ln(\ln N)})\,,\quad\hbox{ et }\Lambda_N=o(N^{1/4})
$$
lorsque $N\to+\infty$. On pose $r=\tfrac12\sqrt{\gamma\lambda^3/N}$ et
\begin{equation}\label{CondInLinBBGKY}
F_N^{in}(X_N,V_N):=\mathcal{Z}_{N,\beta}^{-1}\phi_N^{in}(x_1,v_1)\prod_{j=1}^NM_\beta(v_j)\,,
\end{equation}
pour tout $(x_j,v_j)\in\mathbf{R}^3\times\mathbf{T}^3_\lambda$ 
vérifiant\footnote{Pour $x,y\in\mathbf{T}^3_\lambda$, on pose $|x-y|:=\inf\{|\xi-\eta|\hbox{ t.q. }\xi,\eta\in\mathbf{R}^3\hbox{ avec }\xi=x\hbox{ et }\eta=y\hbox{ mod. }\lambda\mathbf{Z}^3\}$.} $|x_i-x_j|>2r$ pour $1\le i<j\le N$, avec
$$
\mathcal{Z}_{N,\beta}:=\int_{(\mathbf{T}^3_\lambda)^N\times(\mathbf{R}^3)^N}\phi_N^{in}(x_1,v_1)\prod_{j=1}^kM_\beta(v_j)\prod_{1\le i<j\le N}\mathbf{1}_{|x_i-x_j|>2r}dX_NdV_N\,.
$$
(La notation $M_\beta$ a été définie à la formule (\ref{DefMb}).) Pour tout $t\in\mathbf{R}$, on pose\footnote{On identifie les fonctions définies p.p. sur $(\mathbf{T}^3_\lambda)^n$ avec des fonctions de période $\lambda$ en chaque 
variable définies p.p. sur $(\mathbf{R}^3)^n$.}
\begin{equation}\label{EvolLinFN}
F_N(t,X_N,V_N):=F_N^{in}(S^{N,r}_{-t}(X_N,V_N))\quad\hbox{ pour presque tout }(X_N,V_N)\in\Gamma^r_N\,,
\end{equation}
où on rappelle que $S^{N,r}_t$ est le groupe à un paramètre engendré par les équations de Newton (\ref{Newton})-(\ref{Collxk})-(\ref{Collvkvj}) pour un système de $N$ boules de rayon $r$ interagissant au cours de collisions élastiques. 
Soit d'autre part $g_N$ la solution de l'équation de Boltzmann linéaire (\ref{EqBoltzLin}) vérifiant la condition initiale 
$$
g_N(0,x,v)=\lambda^3\left(\int_{\mathbf{T}^3_\lambda\times\mathbf{R}^3}\phi_N^{in}(x,v)M_\beta(v)dxdv\right)^{-1}\phi_N^{in}(x,v)M_\beta(v)\,.
$$
Alors il existe $C_\beta>0$ tel que, pour tout $t_N=o(\sqrt{\ln(\ln N)}/\mu_N)$, l'on ait
$$
\|\lambda^3F_{N:1}-g_N\|_{L^\infty([0,t_N]\times\mathbf{T}^3_\lambda\times\mathbf{R}^3)}\le C_\beta\left(\frac{\mu_N}{(\ln(\ln N))^2}\right)t_N^4\,,
$$
uniformément en $\lambda\ge 1$.
\end{theo}

L'adaptation de la stratégie de Lanford au cas de l'équation de Boltzmann linéaire avait été esquissée dans \cite{LeSp}, mais cette stratégie ne permettait de justifier la limite de Boltzmann-Grad dans ce cadre que sur le même intervalle
de temps que pour l'équation de Boltzmann non linéaire. La nouveauté et l'intérêt du résultat ci-dessus résident dans le fait qu'il démontre la validité de l'équation de Boltzmann non linéaire sur un intervalle de temps qui tend vers l'infini
avec le nombre $N$ de particules.

\subsection{De la mécanique de Newton à l'équation de diffusion}


En utilisant conjointement le théorème \ref{T-BGSR} et le théorème \ref{T-ApproxDiff}, Bodineau, Gallagher et Saint-Raymond démontrent le résultat suivant.

\begin{theo}\label{T-BGSR2}
Soit $\chi\in C^\infty(\mathbf{R}^3)$ à support dans la boule unité $B(0,1)$ telle que
$$
\chi\ge 0\,,\quad\hbox{ et }\int_{\mathbf{R}^3}\chi(y)dy=1\,.
$$
Soit $\zeta\in]0,\tfrac1{12}[$. Posons 
$$
\rho^{in}_N(x):=\lambda_N^{3\zeta}\chi\left(\frac{x}{\lambda^{1-\zeta}}\right)\,,\quad x\in\mathbf{T}^3_{\lambda_N}\,,
$$
en supposant que $\lambda_N=o((\ln(\ln N))^{1/5})$. 

Posons $r=\tfrac12\sqrt{\gamma\lambda_N^3/N}$ et définissons $F_N^{in}$ par la formule (\ref{CondInLinBBGKY}) avec $\phi_N^{in}(x,v)=\rho_N^{in}(x)$ pour tout $(x,v)\in\mathbf{T}^3_{\lambda_N}\times\mathbf{R}^3$. Soit enfin $F_N$ 
définie par la formule (\ref{EvolLinFN}). Alors la suite de fonctions $f_N$ définie par
$$
f_N(\tau,y,v):=\lambda_N^3F_{N:1}(\lambda_N^2\tau,\lambda_Ny,v)
$$
converge étroitement vers la solution de l'équation de diffusion
$$
\partial_t\rho-D\Delta_x\rho=0\,,\qquad y\in\mathbf{T}^3\,,\qquad\qquad \rho\big|_{t=0}=\delta_0\,,
$$
où $D$ est le réel positif défini par la formule (\ref{CoeffDiff}).
\end{theo}

\smallskip
On sait que la solution du problème de Cauchy pour l'équation de la chaleur
$$
\partial_tu-D\partial^2_xu=0\,,\quad (t,x)\in\mathbf{R}_+^*\times\mathbf{R}\,,\qquad u\big|_{t=0}=u^{in}\,,
$$
est donnée par la formule
$$
u(t,x)=\mathbf{E}(u^{in}(x+B_{2Dt}))
$$
où $B_t$ est le mouvement brownien standard dans $\mathbf{R}^3$. En travaillant un peu plus, on peut formuler le théorème \ref{T-BGSR2} en termes de processus stochastiques, comme suit. Notons $\xi_N(t,X_N,V_N)$ la position 
de la particule blanche à l'instant $t$, c'est-à-dire la première composante de $S^{N,r}_t(X_N,V_N)$, puis posons 
$$
\eta_N(t,X_N,V_N):=\lambda_N^{-1}(\xi_N(2\lambda_N^2Dt,X_N,V_N)-x_1)\,.
$$
Soit alors $W_N$ la mesure image de $m_N$ par l'application $(X_N,V_N)\mapsto \eta_N(\cdot,X_N,V_N)\Big|_{\mathbf{R}_+}$ qui associe à $(X_N,V_N)$ le chemin continu $t\mapsto\eta_N(t,X_N,V_N)$ issu de l'origine pour $t=0$ et 
restreint aux temps positifs ou nuls. Alors la mesure $W_N$ (définie sur l'espace des chemins continus tracés dans $\mathbf{R}^3$ et issus de l'origine pour $t=0$) converge étroitement vers la mesure de Wiener. Le lecteur intéressé par 
ce point de vue est invité à en lire une présentation détaillée dans la section 6.2 de \cite{BGSR}.

\subsection{Résultats antérieurs}


Outre l'article originel \cite{La}, on trouve une présentation du théorème de Lanford dans \cite{CIP,CGP} avec des degrés variés de précision sur la mise en \oe uvre de la stratégie de Lanford. La monographie \cite{GSRT} est, à notre 
connaissance, la première référence où la vérification de tous les points techniques de la démonstration soit écrite complètement. 

L'article \cite{BGSR} est la première référence où la validité de l'équation de Boltzmann linéaire soit établie sur des intervalles de temps tendant vers l'infini avec le nombre de particules, et où le mouvement brownien soit obtenu comme 
limite de la dynamique newtonienne déterministe de particules identiques en interaction. L'équation de Boltzmann linéaire avait été établie auparavant dans un autre contexte, celui du {\it gaz de Lorentz} \cite{Lor}. Dans ce modèle, les 
particules noires ne sont pas identiques à la particule blanche, mais au contraire {\it de vitesse initiale nulle et de masse infinie}, de sorte qu'on peut les considérer comme rigoureusement immobiles. Lorsque la distribution des particules 
noires est aléatoire et donnée par un processus ponctuel de Poisson, la fonction de distribution de la particule blanche obéit à une équation de Boltzmann linéaire semblable (mais non exactement identique) à (\ref{EqBoltzLin}). Ce résultat 
a été établi par Gallavotti en 1972 (voir \cite{Gallav}, appendice 1.A.2). Le cas d'une distribution {\it périodique} de particules noires ne peut pas être décrit par une équation de type Boltzmann linéaire \cite{BGW,Gol}, mais nécessite 
une description par un modèle cinétique dans un espace des phases plus gros que l'espace des phases d'une particule ponctuelle \cite{CG,MarStro}. Toutefois, la limite du gaz de Lorentz périodique (sous une hypothèse géométrique 
supplémentaire dite d'{\it horizon fini}) vers la diffusion a été annoncée dans \cite{BuSi}. La stratégie de preuve de \cite{BuSi} utilise un arsenal particulièrement technique de méthodes venant de la théorie ergodique ramenant le problème 
à un codage par une dynamique symbolique analogue à celle du flot géodésique sur les surfaces à courbure négative \cite{Pan}.

\section{Les stratégies de démonstration}


Depuis l'article \cite{La}, les stratégies de démonstration des résultats ci-dessus suivent toutes plus ou moins les étapes que nous allons décrire.

\smallskip
\noindent
\textit{Etape 1.} On exprime la $k$-ième marginale $F_{N:k}$ de la fonction de distributions à $N$ corps en appliquant la formule de Duhamel itérée à la $k$-ième équation de la hiérarchie BBGKY; on obtient ainsi une expression de 
$F_{N:k}$ comme somme {\it finie} de $N-k$ termes, chacun dépendant de manière {\it explicite} de la donnée initiale $F^{in}_N$.

\noindent
\textit{Etape 2.} On passe à la limite dans l'expression de $F_{N:k}$ obtenue à l'étape précédente lorsque $N\to+\infty$ sachant que $r$ est lié à $N$ par la loi d'échelle de Boltzmann-Grad (\ref{BGrad}); on obtient ainsi une série --- 
c'est-à-dire une somme \textit{infinie}, contrairement à l'expression de $F_{N:k}$ obtenue dans l'étape précédente --- qui n'est rien d'autre que l'expression de $F_k$, autrement dit de la $k$-ième inconnue dans la hiérarchie de Boltzmann. 
A nouveau, chaque terme de cette série ne fait intervenir que la suite de données initiales $(F_k\Big|_{t=0})_{k\ge 1}$, qui est bien évidemment connue. 

\noindent
\textit{Etape 3.} La série obtenue à l'étape précédente fournit donc, pour tout $k\ge 1$, l'expression \textit{explicite} de $F_k(t,X_k,V_k)$ pour tout $t$ appartenant à l'intervalle de temps $[0,T^*]$  sur lequel la limite de Boltzmann-Grad peut 
être établie, pour (presque) tout $k$-uplet de positions et de vitesses $(X_k,V_k)\in(\mathbf{R}^3)^k\times(\mathbf{R}^3)^k$. Lorsque $k=1$, on {\og reconnaît\fg} dans cette formule l'expression {\og explicite\fg} de la solution de l'équation 
de Boltzmann comme série de Duhamel.

\smallskip
Un aspect quelque peu déroutant de cette stratégie est que l'on déduit l'équation de Boltzmann de sa solution! Quoiqu'il eût été sans doute plus naturel de déduire l'équation de Boltzmann par un raisonnement portant sur les {\it équations} 
des hiérarchies BBGKY et de Boltzmann et non sur une formule explicite donnant leurs solutions, il semble qu'il y ait une obstruction sérieuse à cela: voir \cite{CIP} pp. 74-75.

\subsection{L'expression donnant $F_{N:k}$}


Notons par abus $F_{N:k}(t)$ la fonction $(X_k,V_k)\mapsto F_{N:k}(t,X_k,V_k)$. La formule de Duhamel appliquée à l'équation pour $F_{N:k}$ donne donc
$$
F_{N:k}(t)=\tilde S^{k,r}_tF^{in}_{N:k}+\int_0^t\tilde S^{k,r}_{t-t_1}\mathbf{C}^{k+1}_NF_{N:k+1}(t_1)dt_1\,,
$$
où 
\begin{equation}\label{DefbCk+1N}
\mathbf{C}^{k+1}_N:=\sum_{j=1}^k\mathcal{C}^{j,k+1}_N\,.
\end{equation}
En procédant de même pour $F_{N:k+1}$, on trouve que
$$
F_{N:k+1}(t_1)=\tilde S^{k+1,r}_{t_1}F^{in}_{N:k+1}+\int_0^{t_1}\tilde S^{k+1,r}_{t_1-t_2}\mathbf{C}^{k+2}_NF_{N:k+2}(t_2)dt_2\,,
$$
expression que l'on reporte dans celle donnant $F_{N:k}$, pour trouver
$$
\begin{aligned}
F_{N:k}(t)=\tilde S^{k,r}_tF^{in}_{N:k}&+\int_0^t\tilde S^{k,r}_{t-t_1}\mathbf{C}^{k+1}_N\tilde S^{k+1,r}_{t_1}F^{in}_{N:k+1}dt_1
\\
&+\int_0^t\int_0^{t_1}\tilde S^{k,r}_{t-t_1}\mathbf{C}^{k+1}_N\tilde S^{k+1,r}_{t_1-t_2}\mathbf{C}^{k+2}_NF_{N:k+2}(t_2)dt_2dt_1\,.
\end{aligned}
$$
En itérant cette procédure, on aboutit à la formule suivante:
\begin{equation}\label{SerieFNk}
\begin{aligned}
F_{N:k}(t)=&\tilde S^{k,r}_tF^{in}_{N:k}+\sum_{j=1}^{N-k}\int_{0\le t_j\le\ldots\le t_1\le t}\tilde S^{k,r}_{t-t_1}\mathbf{C}^{k+1}_N\tilde S^{k+1,r}_{t_1-t_2}\ldots\mathbf{C}^{k+j}_N\tilde S^{k+j,r}_{t_j}F^{in}_{N:k+j}dt_j\ldots dt_1\,.
\end{aligned}
\end{equation}

\subsection{L'expression donnant $F_k$}


En procédant de même, on aboutit à la formule suivante pour $F_k$ apparaissant dans la $k$-ième équation de la hiérarchie de Boltzmann --- toujours en convenant de noter $F_k(t)$ la fonction $(X_k,V_k)\mapsto F_k(t,X_k,V_k)$:
\begin{equation}\label{SerieFk}
F_k(t)=\tilde S^k_tF^{in}_k+\sum_{j\ge 1}\int_{0\le t_j\le\ldots\le t_1\le t}\tilde S^k_{t-t_1}\mathbf{C}^{k+1}\tilde S^{k+1}_{t_1-t_2}\ldots\mathbf{C}^{k+j}\tilde S^{k+j}_{t_j}F^{in}_{k+j}dt_j\ldots dt_1\,,
\end{equation}
où l'on a posé
\begin{equation}\label{DefbCk+1}
\mathbf{C}^{k+1}:=\sum_{j=1}^k\mathcal{C}^{j,k+1}\,,
\end{equation}
et
$$
\tilde S^k_t\Phi(X_k,V_k):=\Phi(X_k-tV_k,V_k)\,.
$$

Il faut donc maintenant comprendre sur les expressions (\ref{SerieFNk}) et (\ref{SerieFk}) comment on peut démontrer la convergence $F_{N:k}\to F_k$ pour tout $k\ge 1$ fixé lorsque $N\to+\infty$. Il s'agit évidemment d'un problème de
passage à la limite dans une série. Il faut donc

\smallskip
\noindent
(a) passer à la limite terme à terme dans la série (\ref{SerieFNk}), c'est-à-dire montrer que
$$
\begin{aligned}
\int_{0\le t_j\le\ldots\le t_1\le t}\tilde S^{k,r}_{t-t_1}\mathbf{C}^{k+1}_N\tilde S^{k+1,r}_{t_1-t_2}\ldots\mathbf{C}^{k+j}_N\tilde S^{k+j,r}_{t_j}F^{in}_{N:k+j}dt_j\ldots dt_1
\\
\to
\int_{0\le t_j\le\ldots\le t_1\le t}\tilde S^k_{t-t_1}\mathbf{C}^{k+1}\tilde S^{k+1}_{t_1-t_2}\ldots\mathbf{C}^{k+j}\tilde S^{k+j}_{t_j}F^{in}_{k+j}dt_j\ldots dt_1
\end{aligned}
$$
pour tout $j\ge 1$ lorsque $N\to+\infty$;

\noindent
(b) construire une série majorante permettant d'appliquer le théorème de convergence dominée afin de justifier que
$$
\begin{aligned}
\sum_{j=1}^{N-k}\int_{0\le t_j\le\ldots\le t_1\le t}\tilde S^{k,r}_{t-t_1}\mathbf{C}^{k+1}_N\tilde S^{k+1,r}_{t_1-t_2}\ldots\mathbf{C}^{k+j}_N\tilde S^{k+j,r}_{t_j}F^{in}_{N:k+j}dt_j\ldots dt_1&
\\
\to\sum_{j=1}^\infty\int_{0\le t_j\le\ldots\le t_1\le t}\tilde S^k_{t-t_1}\mathbf{C}^{k+1}\tilde S^{k+1}_{t_1-t_2}\ldots\mathbf{C}^{k+j}\tilde S^{k+j}_{t_j}F^{in}_{k+j}dt_j\ldots dt_1&
\end{aligned}
$$
en un sens à préciser lorsque $N\to+\infty$ pour tout $k\ge 1$. Voir \cite{CIP}, section 4.4.

\subsection{L'estimation de type Cauchy-Kowalevski}


Curieusement, l'obtention de la majoration requise au (b) est l'étape la moins technique de tout le programme. Cette majoration ressortit à la version abstraite du théorème de Cauchy-Kowalevski de Nirenberg \cite{Nir} et Ovsyannikov 
\cite{Ov}. L'analogie entre les résultats de type Cauchy-Kowalevski et le contrôle des hiérarchies BBGKY est une observation due à Ukai \cite{Uk} qui permet un traitement particulièrement élégant de cette partie du programme.

Pour toute suite $(g_n)_{n\ge 1}$ où $g_n$ est une fonction mesurable définie p.p. sur l'espace des phases à $n$ particules $(\mathbf{R}^3)^n\times(\mathbf{R}^3)^n$, on pose
$$
|g_n|_{n,\beta}:=\supess_{X_n,V_n\in(\mathbf{R}^3)^N}\left(|g_n(X_n,V_n)|\exp(\tfrac12\beta|V_n|^2)\right)\quad\hbox{ avec }|V_n|^2=\sum_{j=1}^n|v_j|^2\,,
$$
et
$$
\|(g_n)_{n\ge 1}\|_{\mu,\beta}:=\sup_{n\ge 1}(|g_n|_{n,\beta}e^{\mu n})\,.
$$
On démontre alors (Proposition 5.3.1 de \cite{GSRT}) que
$$
|\mathbf{C}^{k+1}g_{k+1}(X_k,V_k)|\le C^*\beta^{-3/2}(k\beta^{-1/2}+|V_k|^2)\exp(-\tfrac12\beta|V_k|^2)|g_{k+1}|_{k+1,\beta}
$$
et de même (Proposition 5.3.2 de \cite{GSRT}) que
$$
|\mathbf{C}^{k+1}_Ng_{k+1}(X_k,V_k)|\le C^*\beta^{-3/2}(k\beta^{-1/2}+|V_k|^2)\exp(-\tfrac12\beta|V_k|^2)|g_{k+1}|_{k+1,\beta}
$$
pour tout $g_{k+1}$ p.p. nulle sur le complémentaire de $\Gamma^r_N$, sous l'hypothèse que la loi d'échelle de Boltzmann-Grad (\ref{BGrad}) est vérifiée. Ces inégalités correspondent à l'estimation avec perte de type Cauchy-Kowalevski
suivante (formule (5.3.2) dans \cite{GSRT}):
$$
\|(\mathbf{C}^{k+1}g_k)_{k\ge 1}\|_{\mu',\beta'}\le C^*\left(1+\frac1{\sqrt{\beta}}\right)\left(\frac1{\beta-\beta'}+\frac1{\mu-\mu'}\right)\|(g_n)_{n\ge 1}\|_{\mu,\beta}\,.
$$
(Cette estimation est analogue à la condition (4) de \cite{Nish}.) En travaillant directement sur les séries donnant les expressions de $F_{N:k}$ (solution de la $k$-ième équation dans la hiérarche de BBGKY) et de $F_k$ (solution de la k-ième 
équation dans la hiérarchie de Boltzmann), on démontre le résultat suivant (théorème 6 et remarque 5.1.5 dans \cite{GSRT}):

\begin{prop}
Soient $\beta_0>0$ et $\mu_0\in\mathbf{R}$, et soit 
$$
T^*:=\frac{C^*e^{\mu_0}}{1+\sqrt{\beta_0}}\max_{0\le\beta\le\beta_0}\beta e^{-\beta}(\beta_0-\beta)^2\,.
$$
Il existe $\theta>0$ tel que $\theta T^*<\tfrac12\beta_0$ et tel que, pour tout $(F^{in}_k)_{k\ge 1}$  vérifiant 
$$
\|(F^{in}_k)_{k\ge 1}\|_{\mu_0,\beta_0}<\infty
$$
la hiérarchie de Boltzmann admette une unique solution $(F_k)_{k\ge 1}$ définie pour $t\in[0,T^*[$, v\'erifiant la condition initiale $F_k\big|_{t=0}=F_k^{in}$ pour tout $k\ge 1$, ainsi que la borne 
$$
\|(F_k(t))_{k\ge 1}\|_{\mu_0-\theta t,\beta_0-\theta t}\le 2\|(F^{in}_k)_{k\ge 1}\|_{\mu_0,\beta_0}\,.
$$
\end{prop}

\subsection{Conditions initiales admissibles}


Une difficulté pas tout à fait anodine tient au fait qu'avant passage à la limite dans l'échelle de Boltzmann-Grad, on ne peut pas supposer que les $N$ molécules de gaz considérées sont mutuellement indépendantes. En effet, ces 
particules sont corrélées par la contrainte que les centres de deux quelconques d'entre elles doivent être distants d'au moins $2r$. Autrement dit, la fonction de distribution $F_N$ d'un système de $N$ boules de rayon $r$ doit
vérifier la condition $F_N(t,X_N,V_N)=0$ pour (presque) tout $X_N$ tel que $|x_i-x_j|<2r$ pour au moins une paire d'indices $\{i,j\}\subset\{1,\ldots,N\}$. Sachant que $F_N$ est, de plus, une densité de probabilité, elle ne peut donc 
en aucun cas être de la forme
$$
F_N(t,X_N,V_N)=\prod_{k=1}^Nf(t,x_k,v_k)\,.
$$
On n'obtient la forme factorisée qu'après passage à la limite $N\to+\infty$.

Evidemment, cette observation vaut pour $t=0$, ce qui contraint quelque peu le choix de la condition initiale pour la fonction de distribution des $N$ molécules de gaz. En pratique, on procède comme suit: soit $f^{in}\equiv f^{in}(x,v)$
mesurable sur $\mathbf{R}^3\times\mathbf{R}^3$ ayant vocation à être la donnée initiale pour l'équation de Boltzmann après passage à la limite de Boltzmann-Grad. On supposera qu'il existe $\beta_0>0$ et $\mu_0\in\mathbf{R}$ tels 
que
$$
\iint_{\mathbf{R}^3\times\mathbf{R}^3}f^{in}(x,v)dxdv=1\,,\quad\hbox{et }0\le f^{in}(x,v)\le e^{-\mu_0-\frac12\beta_0|v|^2}\hbox{ p.p. en }(x,v)\in\mathbf{R}^3\times\mathbf{R}^3\,.
$$
On pose
$$
\mathcal{Z}_N:=\int_{\Gamma^r_N}\prod_{k=1}^Nf^{in}(t,x_k,v_k)dX_NdV_N\,.
$$

\begin{prop}
Sous les hypothèses ci-dessus portant sur $f^{in}$, la fonction de distribution à $N$ particules
$$
F_N^{in}(X_N,V_N):=\mathcal{Z}_N^{-1}\mathbf{1}_{X_N\in\Omega^r_N}\prod_{k=1}^Nf^{in}(x_k,v_k)
$$
vérifie les conditions suivantes

\smallskip
\noindent
(a) d'une part, en rappelant la notation $F^{in}_{N:k}=0$ dès que $k>N$, l'on a
$$
\sup_{N\ge 1}\|(F^{in}_{N:k})_{k\ge 1}\|_{\mu_0,\beta_0}<\infty\,,
$$
(b) d'autre part, lorsque $N\to+\infty$ et que $r$ vérifie (\ref{BGrad})
$$
F^{in}_{N:k}\to(f^{in})^{\otimes k}\hbox{ localement uniformément sur }\Omega^0_k\,.
$$
pour tout $k\ge 1$.
\end{prop}

Voir la section 6.1 de \cite{GSRT} pour plus de détails et pour une démonstration de cet énoncé.

\subsection{La convergence terme à terme des séries de Duhamel}


Cette étape est de loin la plus technique de tout le programme, et nous ne pourrons qu'en donner un aperçu très incomplet. Notons $Z_k:=(X_k,V_k)$ et
$$
\begin{aligned}
f^{(k,j)}_N(t,Z_k)&:=\int_{0\le t_j\le\ldots\le t_1\le t}\tilde S^{k,r}_{t-t_1}\mathbf{C}^{k+1}_N\tilde S^{k+1,r}_{t_1-t_2}\ldots\mathbf{C}^{k+j}_N\tilde S^{k+j,r}_{t_j}F^{in}_{N:k+j}(Z_k)dt_j\ldots dt_1\,,
\\
f^{(k,j)}(t,Z_k)&:=\int_{0\le t_j\le\ldots\le t_1\le t}\tilde S^k_{t-t_1}\mathbf{C}^{k+1}\tilde S^{k+1}_{t_1-t_2}\ldots\mathbf{C}^{k+j}\tilde S^{k+j}_{t_j}F^{in}_{k+j}(Z_k)dt_j\ldots dt_1\,.
\end{aligned}
$$
Il s'agit de montrer que les quantités suivantes, nommées \textit{observables}\footnote{En pratique on ne mesure pas physiquement la fonction de distribution, qui dépend de la position $x$ et de la vitesse $v$; mais il est tout à fait possible 
de mesurer ses moyennes en vitesse en un point $x$. Par exemple, on peut mesurer le champ de pression ou de température du gaz en tout point $x$. C'est pourquoi on nomme {\og observables\fg} les moyennes en vitesse de la fonction 
de distribution, par analogie avec la terminologie utilisée en mécanique quantique.}
$$
\begin{aligned}
I^{k,j}_N(t,X_k):=\int_{(\mathbf{R}^3)^k}\phi_k(V_k)f^{(k,j)}_N(t,Z_k)dV_k\,,
\\
I^{k,j}(t,X_k):=\int_{(\mathbf{R}^3)^k}\phi_k(V_k)f^{(k,j)}_N(t,Z_k)dV_k\,,
\end{aligned}
$$
vérifient
$$
I^{k,j}_N(t,X_k)\to I^{k,j}(t,X_k)
$$
localement uniformément sur $[0,T]\times\Omega^0_k$, pour tout $k\ge 1$ et pour toute fonction test $\phi_k\in C_c((\mathbf{R}^3)^k)$. 

Une première approximation consiste à tronquer l'énergie cinétique totale de toutes les particules mises en jeu dans les observables $I^{k,j}_N$ et $I^{k,j}$. D'autre part, les intégrales définissant $f^{k,j}_N$ et $f^{k,j}$ mettent en jeu
le simplexe 
$$
\mathcal{T}_j(t):=\{(t_1,\ldots,t_j)\in\mathbf{R}^j\hbox{ t.q. }0\le t_j\le\ldots\le t_1\le t\}\,.
$$
Une seconde approximation consiste à restreindre l'intégration au sous-domaine $\mathcal{T}_{j,\delta}(t)$ de $\mathcal{T}_j(t)$ défini par
$$
\mathcal{T}_{j,\delta}(t):=\{(t_1,\ldots,t_j)\in\mathcal{T}_j(t)\hbox{ t.q. }t_{i+1}-t_i>\delta\hbox{ pour }1\le i<j\}\,.
$$  
L'interprétation physique de cette approximation est que l'on se restreint à ne considérer que les configurations de particules donnant lieu à des collisions séparées par des durées $\ge\delta$. Ces deux opérations se soldent par des termes d'erreurs faciles à contrôler (voir les propositions 7.1.1, 7.2.1 et 7.3.1 de \cite{GSRT}).

On définit alors de nouvelles observables 
$$
\begin{aligned}
J^{k,j}_{N,R,\delta}(t,X_k)
&:=\int\!\phi_k(V_k)\!\int_{\mathcal{T}_{j,\delta}(t)}\tilde S^{k,r}_{t-t_1}\mathbf{C}^{k+1}_N\ldots\mathbf{C}^{k+j}_N\tilde S^{k+j,r}_{t_j}(\mathbf{1}_{|V_{k+j}|^2\le R^2}F^{in}_{N:k+j})(Z_k)dT_jdV_k\,,
\\
J{k,j}_{R,\delta}(t,X_k)
&:=\int\phi_k(V_k)\int_{\mathcal{T}_{j,\delta}(t)}\tilde S^{k}_{t-t_1}\mathbf{C}^{k+1}\ldots\mathbf{C}^{k+j}\tilde S^{k+j}_{t_j}(\mathbf{1}_{|V_{k+j}|^2\le R^2}F^{in}_{k+j})(Z_k)dT_jdV_k\,,
\end{aligned}
$$
avec la notation usuelle $T_j:=(t_1,\ldots,t_j)$, observables que l'on décompose à leur tour sous la forme
$$
\begin{aligned}
J^{k,j}_{N,R,\delta}(t,X_k)=\sum_{L,M}(\prod_{i=1}^jl_i)I^{k,j}_{N,R,\delta}(t,L,M,X_k)\,,
\\
J^{k,j}_{R,\delta}(t,X_k)=\sum_{L,M}(\prod_{i=1}^jl_i)I^{k,j}_{R,\delta}(t,L,M,X_k)\,.
\end{aligned}
$$
On utilise ici les notations suivantes:
$$
L:=(l_1,\ldots,l_j)\in\{+1,-1\}^j\,,\quad M:=(m_1,\ldots,m_j)\hbox{ avec }1\le m_i\le k+i-1\,,
$$
et
$$
\begin{aligned}
J^{k,j}_{N,R,\delta}(t,L,M,X_k)
\\
=\int\phi_k(V_k)\int_{\mathcal{T}_{j,\delta}(t)}\tilde S^{k,r}_{t-t_1}\mathcal{C}^{l_1,m_1,k+1}_N\ldots\mathcal{C}^{l_j,m_j,k+j}_N\tilde S^{k+j,r}_{t_j}(\mathbf{1}_{|V_{k+j}|^2\le R^2}F^{in}_{N:k+j})(Z_k)dT_jdV_k\,,
\\
I^{k,j}_{R,\delta}(t,L,M,X_k)
\\
=\int\phi_k(V_k)\int_{\mathcal{T}_{j,\delta}(t)}\tilde S^{k}_{t-t_1}\mathcal{C}^{l_1,m_1,k+1}\ldots\mathcal{C}^{l_j,m_j,k+j}\tilde S^{k+j}_{t_j}(\mathbf{1}_{|V_{k+j}|^2\le R^2}F^{in}_{k+j})(Z_k)dT_jdV_k\,.
\end{aligned}
$$
Dans ces formules
$$
\begin{aligned}
\mathcal{C}^{+1,m,k+1}_N(F_{N:k+1})(t,X_k,V_k)&
\\
:=\int_{\mathbf{R}^3\times\mathbf{S}^2}F_{N:k+1}(t,X_k,x_m+2rn,T_{m,k+1}[n]V_{k+1})((v_{k+1}-v_m)\cdot n)_+dv_{k+1}dn&\,,
\\
\mathcal{C}^{-1,m,k+1}_N(F_{N:k+1})(t,X_k,V_k)&
\\
:=\int_{\mathbf{R}^3\times\mathbf{S}^2}F_{N:k+1}(t,X_k,x_m-2rn,V_{k+1})((v_{k+1}-v_m)\cdot n)_+dv_{k+1}dn&\,,
\end{aligned}
$$
et
$$
\begin{aligned}
\mathcal{C}^{+1,m,k+1}(F_{k+1})(t,X_k,V_k)&
\\
:=\int_{\mathbf{R}^3\times\mathbf{S}^2}F_{k+1}(t,X_k,x_m,T_{m,k+1}[n]V_{k+1})((v_{k+1}-v_m)\cdot n)_+dv_{k+1}dn&\,,
\\
\mathcal{C}^{-1,m,k+1}(F_{k+1})(t,X_k,V_k)&
\\
:=\int_{\mathbf{R}^3\times\mathbf{S}^2}F_{k+1}(t,X_k,x_m,V_{k+1})((v_{k+1}-v_m)\cdot n)_+dv_{k+1}dn&\,.
\end{aligned}
$$

On cherche donc à montrer que $I^{k,j}_{N,R,\delta}(t,L,M,X_k)\to I^{k,j}_{R,\delta}(t,L,M,X_k)$ localement uniformément sur $\mathbf{R}_+\times\Omega^0_N$ lorsque $N\to+\infty$. Ces deux expressions se ressemblent évidemment beaucoup 
à première vue, mais comportent des différences subtiles où réside toute la difficulté du problème.

Observons d'abord que les opérateurs $\tilde S^{k+i}_{t_i-t_{i+1}}$ qui interviennent dans la définition de $I^{k,j}_{R,\delta}(t,L,M,X_k)$ correspondent à la dynamique \textit{libre} de $k+i$ particules ponctuelles qui ne se voient pas --- 
dynamique libre qui s'exprime très simplement par la formule
$$
\tilde S^{k+i}_{t_i-t_{i+1}}f_{k+i}(X_{k+i},V_{k+i})=f_{k+i}(X_{k+i}-(t_i-t_{i+1})V_{k+i},V_{k+i})\,.
$$ 
Au contraire, les opérateurs $\tilde S^{k+i,r}_{t_i-t_{i+1}}$ qui interviennent, eux, dans la définition de $I^{k,j}_{N,R,\delta}(t,L,M,X_k)$ correspondent à la dynamique de $k+i$ particules pouvant entrer en collision entre les instants 
$t_{i+1}$ et $t_i$, et il n'existe évidemment pas pour eux de formule aussi simple que pour $\tilde S^{k+i}_{t_i-t_{i+1}}$. 

Autrement dit, dans les observables limites $I^{k,j}_{R,\delta}(t,L,M,X_k)$, seules les collisions intervenant aux instants $t_j<\ldots<t_1$ avec $(t_1,\ldots,t_j)\in\mathcal{T}_{j,\delta}(t)$ sont prises en compte. Les seules collisions prises
en compte à l'instant $t_i$ sont des collisions entre l'une des $k+i-1$ particules destinées à entrer en collision aux instants ultérieurs $t_{i-1}<\ldots<t_1$ et une $k+i$-ième particule additionelle n'ayant jamais rencontré aucune de
ces $k+i-1$ particules avant l'instant $t_i$. Cette dynamique se résume schématiquement par l'arbre de la figure 1.

\begin{figure}\label{F-BoltzTree}

\begin{center}

\includegraphics[width=7cm]{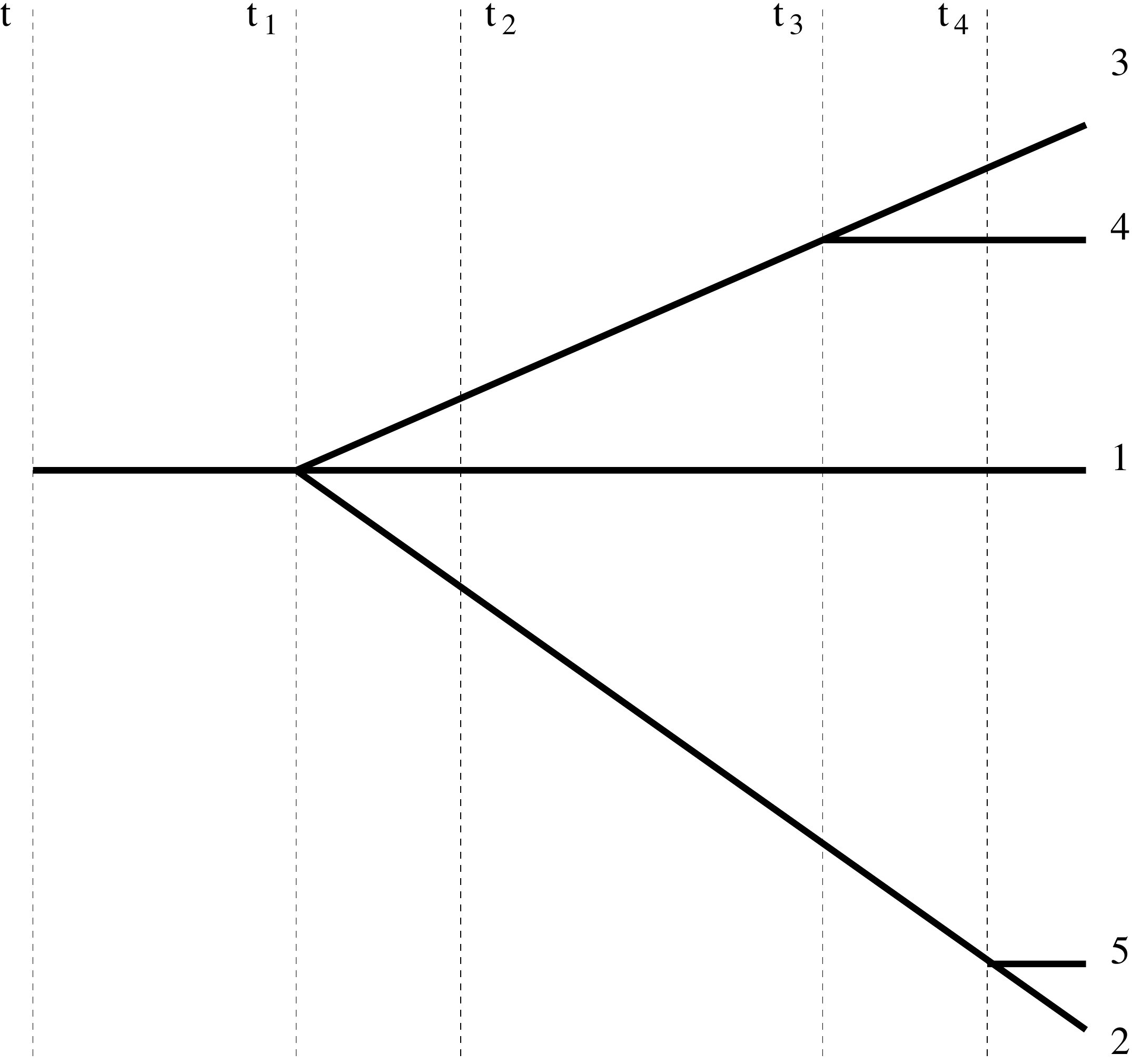}

\caption{Graphe sans recollisions. Ce graphe est un arbre correspondant à une observable du type $I^{1,4}_{R,\delta}(t,L,M,X_1)$, avec $M=(1,1,3,2)$. Cette représentation ne tient pas compte des valeurs de $L=(l_1,l_2,l_3,l_4)\in\{+1,-1\}^4$.}

\end{center}

\end{figure}

Dans les observables $I^{k,j}_{N,R,\delta}(t,L,M,X_k)$ avant passage à la limite de Boltzmann-Grad, on prend évidemment en compte les mêmes collisions que dans $I^{k,j}_{R,\delta}(t,L,M,X_k)$ intervenant aux instants $t_j<\ldots<t_1$.
Et de même que dans $I^{k,j}_{R,\delta}(t,L,M,X_k)$, à l'instant $t_i$, on va prendre en compte les collisions entre l'une des $k+i-1$ particules destinées à entrer en collision aux instants ultérieurs $t_{i-1}<\ldots<t_1$ et une $k+i$-ième 
particule additionelle. La différence est qu'une ou même plusieurs collisions entre les $k+i-1$ particules considérées et la $k+i$-ième qui leur est adjointe à l'instant $t_i$ peuvent parfaitement s'être produites dans l'intervalle de temps 
$]t_{i+1},t_i[$, avant que la $k+i$-ième particule rencontre l'une des $k+i-1$ autres particules à l'instant $t_i$. On désigne de tels évènements sous le nom de \textit{recollisions}. Ces recollisions conduisent à une représentation par des 
graphes comportant des cycles, comme celui de la figure 2.

\begin{figure}\label{F-BBGKYTree}

\begin{center}

\includegraphics[width=7cm]{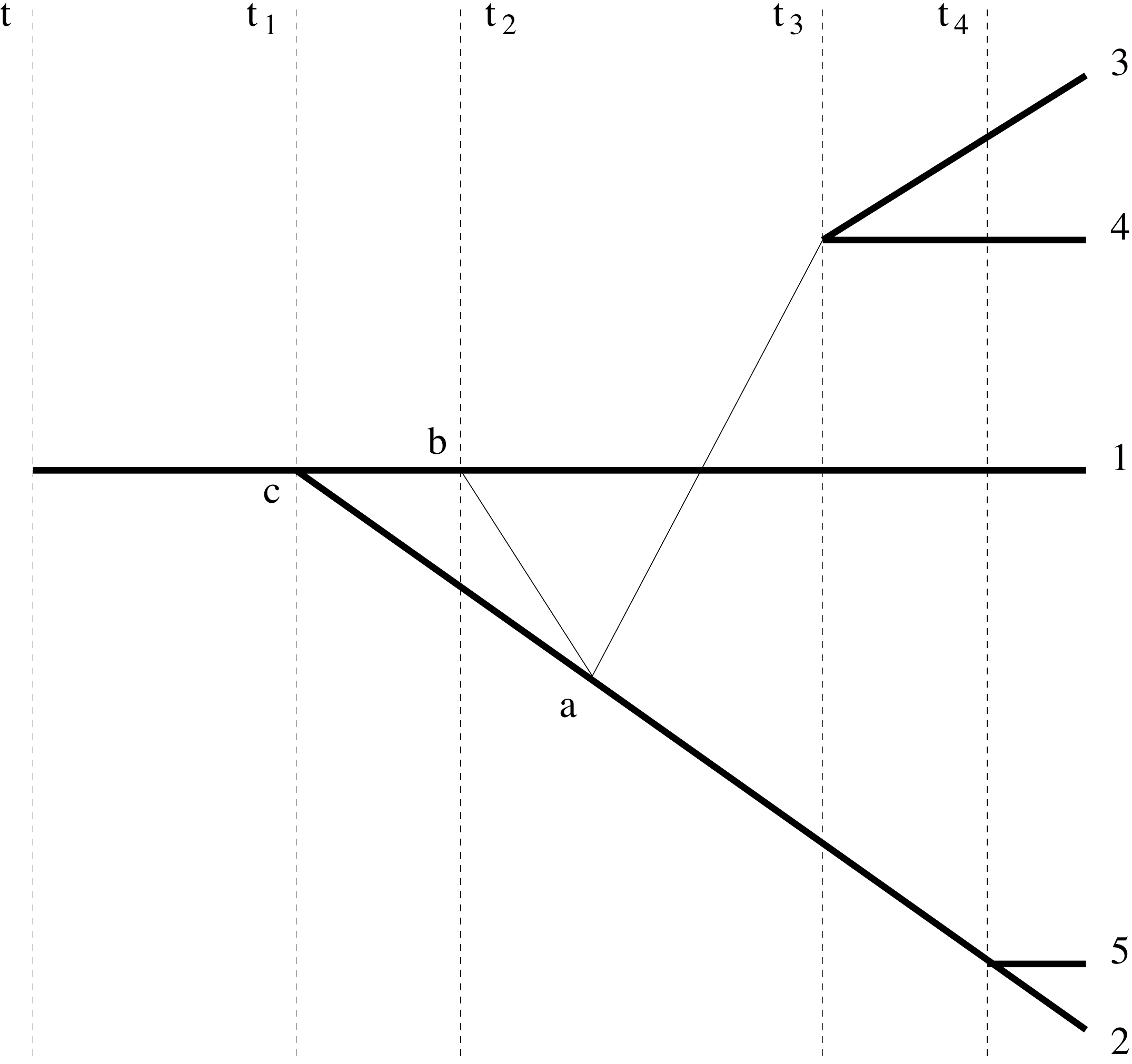}

\caption{Graphe avec recollisions. Comme dans le cas de l'arbre de la figure 1, ce graphe correspond à une observable du type $I^{1,4}_{N,R,\delta}(t,L,M,X_1)$ avec un nombre de particules fini $N$ et pour $M=(1,1,3,2)$. Toutefois, 
avant d'entrer en collision avec la particule n$^\circ$~1 à l'instant $t_2$, la particule n$^\circ$~3 a déjà rencontré la particule n$^\circ$~2 dans l'intervalle de temps $]t_3,t_2[$. Cette recollision est à l'origine du cycle $abca$ sur la figure.}

\end{center}

\end{figure}

Il est bien évident que les recollisions vont à l'encontre de la théorie cinétique et de l'équation de Boltzmann, puisque les recollisions sont précisément des collisions qui ne sont décrites par aucun des opérateurs intégraux du type
$\mathcal{C}^{\pm 1,m_1,k+1}_N$. Autrement dit, il s'agit de collisions que l'on ne peut espérer décrire par l'intégrale des collisions de Boltzmann. Fort heureusement, on démontre que, dans la limite de Boltzmann-Grad, les
recollisions peuvent être négligées. Cela n'est pas très surprenant: rappelons qu'il s'agit de passer à la limite dans $F_{N:k}$ pour tout $k\ge 1$ \textit{fixé} lorsque $N\to+\infty$. Estimer l'importance des recollisions dans l'intervalle 
de temps $]t_{i+1},t_i]$ consiste à comparer la probabilité que $2$ parmi $k+i-1$ particules se rencontrent dans cet intervalle de temps à celle que l'une de ces $k+i-1$ particules rencontre l'une des $N-k-i+1$ autres particules à
un instant $t_i>t_{i+1}$. Comme $N\to+\infty$ tandis que $k,i$ sont fixés et $r\to 0$, il est évident que les événements du premier type (c'est-à-dire les recollisions) sont statistiquement négligeables par rapport aux seconds. Toutefois, la
démonstration précise de ce fait est loin d'être simple: elle occupe plus d'une vingtaine de pages dans \cite{GSRT} (voir chapitres 12-14).

Une deuxième différence entre les intégrales de collision $\mathcal{C}^{\pm 1,m_1,k+1}$ intervenant dans les observables limites $I^{k,j}_{R,\delta}(t,L,M,X_k)$ et $\mathcal{C}^{\pm 1,m_1,k+1}_N$ à $N$ fini intervenant dans 
les observables $I^{k,j}_{N,R,\delta}(t,L,M,X_k)$ est la délocalisation des collisions, autrement dit le fait que l'opérateur $\mathcal{C}^{\pm 1,m_1,k+1}_N$ fait intervenir la restriction de $F_{N:k+1}$ à l'hypersurface d'équation 
$|x_{k+1}-x_j|=2r$, tandis que l'opérateur limite correspondant $\mathcal{C}^{\pm 1,m_1,k+1}$ fait intervenir la restriction de $F_{k+1}$ au sous-espace vectoriel d'équation $x_{k+1}-x_j=0$ qui est, lui, de codimension $3$. Le
point suivant mérite d'être noté: dans la limite de Boltzmann-Grad, on a $r\to 0$, de sorte que les trajectoires $S^{N,r}_t(Z_N)$ et $S^{N,r}_t(Z'_N)$ issues de deux points voisins $Z_N$ et $Z'_N$ de l'espace des phases de $N$ 
particules peuvent diverger très rapidement (exponentiellement) à cause de la courbure du bord de $\Omega^r_N$ qui tend vers l'infini. Il n'en est pas de même avec les trajectoires libres $S^{N}_t(Z_N)$ et $S^{N}_t(Z'_N)$, 
pour lesquelles le calcul explicite montre que $|S^N_t(Z_N)-S^N_t(Z'_N)|=O(t)$.

Pour $f^{in}$ lipschitzienne, on arrive finalement à l'estimation suivante (voir la section 14.2.4 dans \cite{GSRT}):
$$
\begin{aligned}
\left|\int_{(\mathbf{R}^3)^k}\phi(V_k)(F_k-F_{N:k})(t,X_k,V_k)dV_k\right|
\\
\le
C\left(2^{-n}+e^{-C\beta_0R^2}+\frac{n^2\delta}{T}\right)\|\phi\|_{L^\infty}\sup_{N\ge 1}\|(F^{in}_{N:j})_{j\ge 1}\|_{\mu_0,\beta_0}
\\
+
Cn^2(k+n)\left(R\eta^2+R^3\left(\frac{a}{r_0}\right)^2+R\left(\frac{r_0}{\delta}\right)^2\right)\|\phi\|_{L^\infty}\sup_{N\ge 1}\|(F^{in}_{N:j})_{j\ge 1}\|_{\mu_0,\beta_0}
\\
+Cr(k+n)\|\phi\|_{L^\infty}\|(F^{in}_j)_{j\ge 1}\|_{\mu_0,\beta_0}+C\frac{(k+n)^2}{N}\|\phi\|_{L^\infty}\|(F^{in}_j)_{j\ge 1}\|_{\mu_0,\beta_0}
\\
+Cnr\|\nabla_xf^{in}\|_{L^\infty}\|\phi\|_{L^\infty}\|(F^{in}_j)_{j\ge 1}\|_{\mu_0,\beta_0}
\end{aligned}
$$
où on rappelle que $F^{in}_{N:j}=0$ lorsque $j>N$, et où $a,r_0$ et $\eta$ sont des paramètres positifs vérifiant $a\ll r_0\ll \eta\delta$. En choisissant $n=C_1|\ln r|$ et $R=C_2\sqrt{|\ln r|}$ avec $C_1,C_2\gg 1$, puis $\delta=\sqrt{r}$
et enfin $r_0=r^{3/4}$, on trouve que l'erreur ci-dessus est $O(r^{\alpha})$ pour tout $\alpha<\tfrac12$.

\subsection{Le principe du maximum dans le régime linéaire}


On a vu plus haut que, lorsqu'on cherche à passer de la mécanique de Newton à l'équation de Boltzmann linéaire, les hiérarchies BBGKY et de Boltzmann sont rigoureusement identiques à celles utilisées dans la limite de Boltzmann-Grad 
conduisant à l'équation de Boltzmann non linéaire. La seule différence est la choix de la donnée initiale (\ref{CondInLinBBGKY}) au lieu de (\ref{CondInBBGKY}). 

Supposons que la fonction $f_N^{in}$ intervenant dans la condition initiale (\ref{CondInLinBBGKY}) est de la forme $f_N^{in}(x,v)=\phi_N^{in}(x)M_\beta(v)$ avec
$$
\frac1{\mu_N}\le\phi_N^{in}(x,v)\le\mu_N\,,\quad|\nabla f^{in}(x,v)|\le\Lambda_N
$$
pour tout $(x,v)\in\mathbf{T}^3_\lambda\times\mathbf{R}^3$, avec $\mu_N\ll\sqrt{\ln(\ln N)}$ et $\Lambda_N\ll N^{1/4}$. En appliquant le principe du maximum à l'équation de Liouville vérifiée par la fonction de distribution des $N$ particules 
(plus précisément des $N-1$ particules noires et de la particule blanche), on aboutit facilement (voir la proposition 3.2 de \cite{BGSR}) à l'estimation
\begin{equation}\label{EstimLinFNk}
0\le F_{N:k}(t,X_k,V_k)\le\mu_N^2M_{N:k,\beta}\le\frac{\mu_N^2M^{\otimes k}_\beta(V_k)}{(\lambda^3(1-\tfrac43\pi r))^k}\,.
\end{equation}
Dans cette inégalité, on a noté
$$
M_{\beta,N}(X_N,V_N):=\frac{\mathbf{1}_{\Omega^r_N}(X_N)}{\widetilde{\mathcal{Z}_N}}M^{\otimes N}_\beta(V_N)\,,\quad\hbox{ avec }\widetilde{\mathcal{Z}_N}:=\int_{\mathbf{T}^3_\lambda}\mathbf{1}_{\Omega^r_N}(X_N)dX_N\,,
$$
où on rappelle que la notation $M_\beta$ pour la maxwellienne est définie dans (\ref{DefMb}), tandis que
$$
M_{\beta,N:k,}(X_k,V_k):=\int_{(\mathbf{T}^3_\lambda\times\mathbf{R}^3)^{N-k}}M_{\beta,N}(X_N,V_N)dx_{k+1}\ldots dx_Ndv_{k+1}\ldots dv_N\,.
$$

Dans le régime linéaire, on utilisera l'estimation (\ref{EstimLinFNk}) à la place des estimations de type Cauchy-Kowalevski utilisées dans le régime non linéaire. L'intérêt de cette estimation est qu'elle peut être propagée globalement 
en temps, au contraire des estimations de type Cauchy-Kowalevski. Ceci est l'une des raisons qui permettront de justifier l'équation de Boltzmann linéaire sur des plages de temps qui tendent vers l'infini avec le nombre $N$ de 
particules. Comme ce point précis est le progrès majeur atteint dans \cite{BGSR}, et qu'il est essentiel pour arriver à la limite de diffusion (Théorème \ref{T-BGSR2}), le fait de pouvoir utiliser des estimations $L^\infty$ dans la théorie 
de l'équation de Boltzmann linéaire est évidemment crucial.

\subsection{La procédure d'élagage dans le régime linéaire}


Revenons à l'expression de $F_{N:k}$ fournie par (\ref{SerieFNk}), que l'on mettra sous la forme
\begin{equation}\label{ExprLinFNk}
F_{N:k}(t)=\sum_{j=0}^{N-k}Q^{k,k+j}_N(t)F^{in}_{N:k+j}\,,
\end{equation}
où
\begin{equation}\label{QNkj}
\begin{aligned}
Q^{k,k}_N(t)&:=\tilde S^{k,r}_t\,,\qquad\qquad\hbox{ et }
\\
Q^{k,k+j}_N(t)&:=\int_{0\le t_j\le\ldots\le t_1\le t}\tilde S^{k,r}_{t-t_1}\mathbf{C}^{k+1}_N\tilde S^{k+1,r}_{t_1-t_2}\ldots\mathbf{C}^{k+j}_N\tilde S^{k+j,r}_{t_j}dt_j\ldots dt_1\,.
\end{aligned}
\end{equation}

L'interprétation probabiliste de l'équation de Boltzmann linéaire montre qu'elle décrit un régime où les temps de collision entre la particule blanche et l'une des particules noires sont distribués sous une loi exponentielle (voir \cite{Pa})
de paramètre indépendant de la solution (au contraire du cas de l'équation de Boltzmann non linéaire). Par conséquent, il y a en moyenne un nombre fini constant de collisions par unité de temps.

Choisissons $\tau>0$ et une suite $(n_j)_{j\ge 1}$ d'entiers que l'on précisera plus loin. Ecrivons l'expression (\ref{ExprLinFNk}) entre les instants $t-\tau$ et $t$:
\begin{equation}\label{LinFN1n1}
F_{N:1}(t)=\sum_{l=1}^{n_1-1}Q^{1,l}_N(\tau)F_{N:l}(t-\tau)+R_{1,n_1}(t-\tau,t)\,,
\end{equation}
puis, en réitérant la même opération sur l'intervalle de temps $[t-2\tau,t-\tau]$:
\begin{equation}\label{LinFN1n2}
\begin{aligned}
F_{N:1}(t)=\sum_{l_1=1}^{n_1-1}\sum_{l_2=1}^{n_2-1}Q^{1,l_1}_N(\tau)Q^{l_1,l_1+l_2}_N(\tau)F_{N:l_1+l_2}(t-2\tau)&
\\
+R_{1,n_1}(t-\tau,t)+\sum_{l_1=1}^{n_1-1}Q^{1,l_1}_N(\tau)R_{l_1,n_2}(t-2\tau,t-\tau)&\,,
\end{aligned}
\end{equation}
en notant
$$
R_{k,m}(t',t):=\int_{t'}^t\int_{t'}^{t_1}\ldots\int_{t'}^{t_{m-1}}\tilde S^{k,r}_{t-t_1}\mathbf{C}^{k+1}_N\tilde S^{k+1,r}_{t_1-t_2}\mathbf{C}^{k+2}_N\ldots\mathbf{C}^{k+m}_NF_{N:k+m}(t_m)dt_m\ldots dt_1\,.
$$
L'idée est que le terme $R_{k,m}(t',t)$ décrit la contribution à la marginale d'ordre $k$ de $F_N$ des trajectoires comportant au moins $m$ collisions dans l'intervalle de temps $[t',t]$. Compte tenu de l'observation ci-dessus sur
le nombre de collisions par unité de temps, on s'attend à ce que ce terme devienne négligeable lorsque $m$ tend vers l'infini.

Plus généralement, on écrira
\begin{equation}\label{Elag}
F_{N:1}(t)=F^K_{N:1}(t)+R^K_N(t)\,,
\end{equation}
avec
$$
F^K_{N:1}(t):=\sum_{l_1=1}^{n_1-1}\ldots\sum_{l_K=1}^{n_K-1}Q^{1,L_1}_N(\tau)Q^{L_1,L_2}_N(\tau)\ldots Q^{L_{K-1},L_K}_N(\tau)F_{N:L_K}(t-K\tau)
$$
où $L_m:=l_1+\ldots+l_m$, et
$$
R^K_N(t)=\sum_{k=1}^K\sum_{l_1=1}^{n_1-1}\!\ldots\!\!\!\sum_{l_K=1}^{n_{k-1}-1}Q^{1,L_1}_N(\tau)Q^{L_1,L_2}_N(\tau)\ldots Q^{L_{k-2},L_{k-1}}_N(\tau)R_{L_{k-1},n_k}(t\!-\!k\tau,t\!-\!(k\!-\!1)\tau).
$$
En choisissant convenablement la suite $(n_k)_{k\ge 1}$, on s'attend donc à ce que la contribution principale soit le terme $F^K_{N:1}$ et que le terme $R^K_N(t)$ tende vers $0$ (en un sens à préciser) dans l'asymptotique de 
Boltzmann-Grad. En pratique, on prendra $n_k=2^k$; on montre alors (Proposition 4.2 de \cite{BGSR}) que 
$$
\lambda^3\|R^K_N(t)\|_{L^\infty(\mathbf{T}^3_\lambda\times\mathbf{R}^3)}=O(\mu_N^2)\,.
$$
L'utilisation conjointe des estimations $L^\infty$ venant du principe du maximum (\ref{EstimLinFNk}) et la représentation (\ref{Elag}) permet d'aller au-delà du temps $T^*$ fourni par les bornes de type Cauchy-Kowalevski, et d'obtenir
la validité de l'équation de Boltzmann linéaire sur un intervalle de temps très long, tendant vers $0$ avec le nombre $N$ de particules. Le fait de rejeter dans le terme de reste $R^K_N$ les trajectoires comportant trop de collisions
par unité de temps correspond, après élimination des recollisions, à une procédure d'élagage sur les arbres représentant l'histoire de la particule marquée, comme expliqué sur l'exemple de la figure 3.

\begin{figure}\label{F-ElagTree}

\begin{center}

\includegraphics[width=7cm]{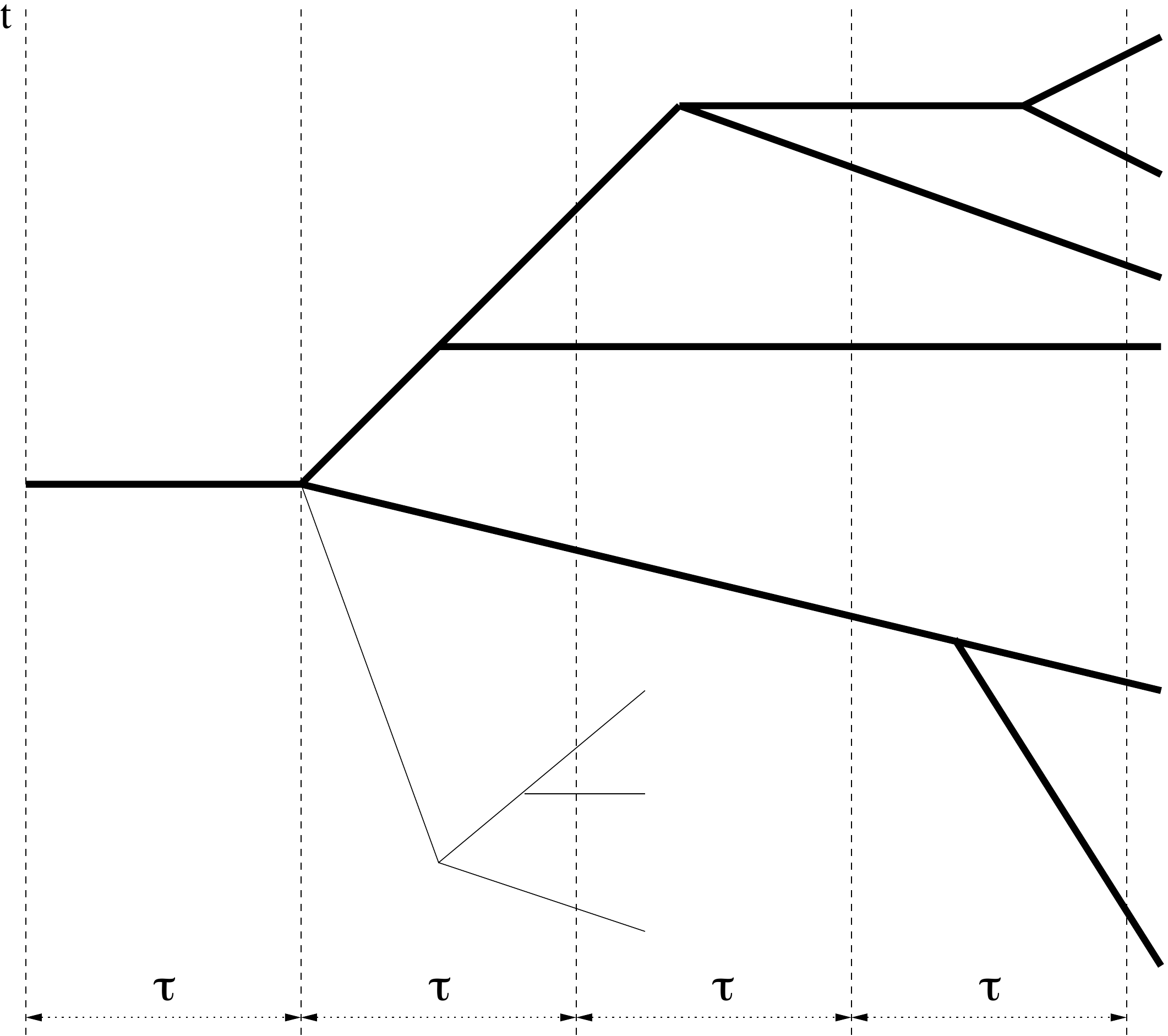}

\caption{En posant $n_k=2^k$, l'arbre en traits gras est retenu puisque le nombre de branches croisant l'axe $t-k\tau$ est inférieur ou égal à $n_k-1$. En revanche, l'arbre complet contenant les branches en traits fins n'est pas retenu
puisqu'il possède $6>n_2=3$ branches croisant l'axe $t-2\tau$.}

\end{center}

\end{figure}


\section{Remarques diverses}


\subsection{Apparition de l'irréversibilité?}


Nous avons déjà évoqué dans l'introduction de cet exposé la question de la réversibilité des équations de Newton et de l'irréversibilité dans l'équation de Boltzmann qui se traduit par le théorème $H$ de Boltzmann. Précisons ces deux
points. La réversibilité mécanique des équations de Newton (\ref{Newton}) signifie exactement que, pour tout $t\in\mathbf{R}$,
\begin{equation}\label{RevMec}
R_N\circ S^{N,r}_t\circ R_N\circ S^{N,r}_t=\hbox{Id}_{\Gamma^r_N}\,,\quad\hbox{ où }R_N(X_N,V_N):=(X_N,-V_N)
\end{equation}
Donc la solution $F_N(t):=\tilde{S}^{N,r}_tF^{in}$ de l'équation de Liouville (\ref{Liouville}) vérifie, pour toute fonction de distribution à $N$ particules $F^{in}_N$ définie (p.p.) sur $\overline{\Gamma^r_N}$ et tout $t\ge 0$, l'identité
$$
\tilde R_N\tilde S^{N,r}_t\tilde R_NF_N(t)=\tilde F^{in}_N\,,\qquad\hbox{ en notant }\tilde R_N\phi_N(X_N,V_N):=\phi_N(X_N,-V_N)\,.
$$

Soit d'autre part $f^{in}$, densité de probabilité sur $\mathbf{R}^3\times\mathbf{R}^3$ vérifiant les hypothèses du théorème \ref{T-Lanford}; notons $S^B_tf^{in}(x,v):=f(t,x,v)$ la solution de l'équation de Boltzmann (\ref{EqBoltz}) de donnée
initiale $f^{in}$. Le théorème H de Boltzmann (voir \cite{CIP}, sections 3.2 et 3.4) assure que la quantité 
$$
H(f(t)):=\iint_{\mathbf{R}^3\times\mathbf{R}^3}f(t,x,v)\ln f(t,x,v)dxdv
$$
satisfait 
\begin{equation}\label{ThmH}
H(f(t))\le H(f^{in})\quad\hbox{ pour tout }0\le t<T^*
\end{equation}
avec inégalité stricte lorsque $f^{in}$ n'est pas une maxwellienne locale, c'est-à-dire lorsque $f^{in}(x,v)$ n'est pas de la forme $f^{in}(x,v)=\rho(x)M_{\beta(x)}(v-u(x))$ avec $\rho(x)\ge 0$ et $\beta(x)>0$ tandis que $u(x)\in\mathbf{R}^3$. 
Donc, si $f^{in}$ vérifie les hypothèses du théorème \ref{T-Lanford} sans être une maxwellienne locale, pour tout $t\in]0,T^*/2[$, l'on a
$$
H(\tilde R_1S^B_t\tilde R_1S^B_tf^{in})=H(S^B_t\tilde R_NS^B_tf^{in})\le H(\tilde R_NS^B_tf^{in})=H(S^B_tf^{in})<H(f^{in})\,,
$$
ce qui montre que 
\begin{equation}\label{IrrB}
\tilde R_1S^B_t\tilde R_1S^B_tf^{in}\not= f^{in}\,,\qquad\hbox{ pour }0<t<T^*/2\,.
\end{equation}
Autrement dit, l'équation de Boltzmann ne vérifie pas la propriété de réversibilité mécanique, au contraire de l'équation de Liouville. Cette conséquence élémentaire du théorème H de Boltzmann est à l'origine de la controverse qui
entoura la théorie cinétique des gaz et ce que Hilbert appelait la {\og conception atomique\fg}. 

Mais la formulation précise de la limite de Boltzmann-Grad donnée par Lanford \cite{La} permet de voir qu'il n'y a aucune contradiction entre la réversibilité mécanique des équations de Newton et le théorème H de Boltzmann. En effet,
la réversibilité mécanique des équations de Newton est une propriété qui s'énonce dans l'espace des phases $\Gamma^r_N$ à $N$ particules, ou, de façon équivalente, sur la fonction de distribution jointe $F_N$ des $N$ particules
qui est solution de l'équation de Liouville. L'équation de Boltzmann, elle, est vérifiée par la première marginale $F_{N:1}$ de $F_N$ dans la limite de Boltzmann-Grad. Lorsqu'on passe de $F_N$ à sa première marginale, on perd 
évidemment beaucoup d'information sur le système des $N$ particules, et cette perte d'information contribue à l'inégalité stricte dans le théorème $H$ au niveau de l'équation de Boltzmann. Ainsi, la tranformation $\tilde R_1$ dans
(\ref{IrrB}) consiste à effectuer la substitution $v\mapsto -v$ sur {\it une} particule typique, ce qui n'est évidemment pas la même chose que d'effectuer la même transformation {\it simultanément sur toutes} les $N$ particules ---
ce qui est justement l'effet de la transformation $R_N$ intervenant dans la définition de la réversibilité mécanique des équations de Newton. 

Un autre facteur contribuant à l'irréversibilité est le fait que la limite de Boltzmann-Grad implique que $r\to 0$. Tant que $r>0$, les lois de collision (\ref{Collxk})-(\ref{Collvkvj}) sont réversibles car le vecteur $n_{kl}$ est déterminé de 
manière unique par la position des particules n$^\circ$~$k$ et $l$. En revanche, lorsque $r\to 0$, la définition de l'intégrale de collision $\mathcal{C}(f)$  dans (\ref{IntColl}) met en jeu la relation (\ref{Collvv*v'v'*}), où le vecteur $n$ 
analogue à $n_{kl}$ est cette fois aléatoire et distribué de manière uniforme sur la sphère. 

Enfin, la condition initiale joue évidemment un rôle essentiel dans l'irréversibilité. Comme observé dans l'appendice de \cite{LeSp}, en supposant $F^{in}_N$ de la forme (\ref{CondInBBGKY}), la fonction de distribution à $N$ particules 
après inversion des vitesses à l'instant $t$, soit $R_N\tilde S^{N,r}_tF^{in}_N$ n'est en général pas de la forme (\ref{CondInBBGKY}) --- ni même d'une forme qui permettrait d'appliquer le théorème \ref{T-Lanford}. On ne peut donc pas
décrire l'évolution ultérieure des $N$ particules après inversion des vitesses à l'instant $t$ en utilisant l'équation de Boltzmann, même dans la limite de Boltzmann-Grad.

Peut-on dès lors parler d'apparition de l'irréversibilité dans la limite de Boltzmann-Grad comme d'un principe fondamental de la physique s'appliquant au cas particulier de la dynamique des gaz? Ces quelques remarques montrent
qu'il serait plus juste de dire que l'irréversibilité réside dans la nature même des objets mathématiques et de la limite asymptotique que l'on considère.

On trouvera une discussion très complète des aspects techniques de cette question dans \cite{CIP}, sections 3.6 et 4.7. L'appendice de \cite{LeSp} met en lumière le rôle de la condition initiale dans l'apparition de l'irréversibilité au 
niveau de la hiérarchie de Boltzmann (et pas seulement de l'équation de Boltzmann). Le lecteur intéressé par les aspects historiques de cette même question est renvoyé au chapitre 5 de \cite{CeLB} (voir également les sections 4.1, 
7.5 et 7.6, ainsi que les chapitres 6 et 11, {\it ibid.}).

\subsection{Le cas des potentiels à courte portée}


Dans cet exposé, on s'est volontairement limité à décrire la limite de Boltzmann-Grad dans le seul cas où l'interaction entre molécules de gaz correspond à des collisions élastiques entre sphères dures. Cette hypothèse n'est évidemment
pas très réaliste du point de vue physique. En fait, un énoncé analogue au théorème \ref{T-Lanford} est démontré complètement dans l'ouvrage \cite{GSRT} (partie III) lorsque l'interaction entre molécules est définie par un potentiel répulsif 
radial de la forme $x\mapsto\Phi(|x|)$, où $\Phi\in C^2(\mathbf{R}_+^*)$ vérifie les conditions suivantes:
$$
\lim_{R\to 0^+}\Phi(R)=+\infty\,,\quad\Phi'(R)<0\hbox{ et }(R^2\Phi'(R))'\ge 0\hbox{ sur }]0,1[\,,\quad\Phi(R)=0\hbox{ sur }[1,+\infty[\,.
$$
(La troisième condition garantit la monotonie de l'angle de déflection en fonction du paramètre d'impact lors d'une collision: voir le lemme 8.3.1 dans \cite{GSRT} et l'appendice de \cite{PSS}, notamment la formule (A8).) On trouvera dans
\cite{PSS}, notamment dans la section 8, une discussion du cas de potentiels radiaux non monotones (comme par exemple un potentiel de type Lennard-Jones tronqué).

Les équations de Newton (\ref{Newton}) sont alors remplacées par le système hamiltonien
$$
\frac{dx_k}{dt}(t)=v_k(t)\,,\quad\frac{dv_k}{dt}(t)=-\frac1{r}\sum_{l=1\atop l\not=k}^N\Phi'\left(\frac{|x_l(t)-x_k(t)|}{r}\right)\frac{x_l(t)-x_k(t)}{|x_l(t)-x_k(t)|}\,.
$$
On trouve dans la thèse de King \cite{King} (malheureusement jamais publiée) comment adapter la stratégie de Lanford au cas de tels potentiels d'interaction --- là encore, les mêmes points techniques que dans \cite{La} restaient à 
vérifier. Notons que, dans ce cas, la définition des marginales doit être légèrement modifiée (voir \cite{So}, Appendice A1, et \cite{GSRT}, chapitre 9). La condition de support compact sur le potentiel est absolument essentielle 
pour garantir que la section efficace de collision intervenant dans l'intégrale de collision (\ref{IntColl}) est intégrable par rapport à la variable angulaire $n$. Dans le cas des sphères dures, cette section efficace est proportionnelle à 
$|\cos(v-v_*,n)|$. Même en supposant que $\Phi$ est une fonction à décroissance rapide ou exponentielle, mais non identiquement nulle (ou constante) à l'infini, la présence de \textit{collisions rasantes} (c'est-à-dire de collisions 
où la vitesse relative des particules est déviée très faiblement au cours de la collision) fait diverger l'intégrale en angle de la section efficace. Cette circonstance modifie  de façon radicale la théorie mathématique de l'équation de 
Boltzmann. Sa validité n'est à ce jour pas démontrée à partir des équations de Newton pour de tels potentiels. 

\bigskip
\textit{Remerciements} -- L'auteur remercie Thierry Bodineau, Isabelle Gallagher et Laure Saint-Raymond pour leurs remarques et leurs suggestions sur le texte de cet exposé.


\end{document}